\documentclass[journal]{IEEEtran}
\usepackage{cite}
\usepackage{graphicx}
\usepackage{amsmath}
\usepackage{times}
\usepackage{latexsym}
\usepackage{graphicx}
\usepackage{bm}
\usepackage{amssymb}
\usepackage[center]{caption2}
\usepackage{stfloats}
\usepackage{cases}
\usepackage{array}
\usepackage{setspace}
\usepackage{fancyhdr}
\usepackage{citesort}
\usepackage{color}

\renewcommand{\QED}{\QEDopen}
\newtheorem{theorem}{Theorem}

\newtheorem{corollary}{Corollary}
\newtheorem{proposition}{Proposition}

\def\proof{\noindent\hspace{2em}{\itshape Proof: }}
\def\endproof{\hspace*{\fill}~$\square$\par\endtrivlist\unskip}

\begin{document}
\title{Outage Probability of Dual-Hop Multiple Antenna AF Systems with Linear Processing in the Presence of Co-Channel Interference}
\author{Guangxu Zhu, Caijun Zhong,~\IEEEmembership{Member,~IEEE,} Himal A. Suraweera,~\IEEEmembership{Member,~IEEE,} Zhaoyang Zhang,~\IEEEmembership{Member,~IEEE} and Chau Yuen~\IEEEmembership{Member,~IEEE}
%\thanks{Manuscript received May 6, 2008; revised September 09, 2008 and 19 January 2009; accepted 30 March, 2009. The associate
%        editor coordinating the review of this manuscript and approving it for
%        publication was Prof. Gerald Matz. The work was supported by EPSRC under grant EP/D058716/1. This paper was presented in part at the IEEE
%        International Symposium on Information Theory, Toronto, Canada, July 2008.}
\thanks{Guangxu Zhu, Caijun Zhong and Zhaoyang Zhang are with the Institute of Information and Communication Engineering, Zhejiang University, China. (email:caijunzhong@zju.edu.cn).}
\thanks{Himal A. Suraweera is with the Department of Electrical \& Electronic
Engineering, University of Peradeniya, Peradeniya 20400, Sri Lanka (email:
himal@ee.pdn.ac.lk)}
\thanks{Chau Yuen is with the Singapore University of Technology and Design, 20 Dover
Drive, Singapore 138682 (email: yuenchau@sutd.edu.sg)}}

\markboth{IEEE TRANSACTIONS ON Wireless Communications,~Vol.~x, No.~xx,~xx~201x} {IEEE TRANSACTIONS ON COMMUNICATIONS,~Vol.~x,
No.~xx,~xx~201x}

\maketitle

\begin{abstract}
This paper considers a dual-hop amplify-and-forward (AF) relaying
system where the relay is equipped with multiple antennas, while the source and the destination are equipped with a single antenna. Assuming that the relay is subjected to co-channel interference (CCI) and additive white Gaussian noise (AWGN) while the destination is corrupted by AWGN only, we propose three heuristic relay precoding schemes to combat the CCI, namely, 1) Maximum ratio combining/maximal ratio transmission (MRC/MRT), 2) Zero-forcing/MRT (ZF/MRT), 3) Minimum mean-square error/MRT (MMSE/MRT). We derive new exact outage expressions as well as simple high signal-to-noise ratio (SNR) outage approximations for all three schemes. Our findings suggest that both the MRC/MRT and the MMSE/MRT schemes achieve a full diversity of $N$, while the ZF/MRT scheme achieves a diversity order of $N-M$, where $N$ is the number of relay antennas and $M$ is the number of interferers. In addition, we show that the MMSE/MRT scheme always achieves the best outage performance, and the ZF/MRT scheme outperforms the MRC/MRT scheme in the low SNR regime, while becomes inferior to the MRC/MRT scheme in the high SNR regime. {Finally, in the large $N$ regime, we show that both the ZF/MRT and MMSE/MRT schemes are capable of completely eliminating the CCI, while perfect interference cancelation is not possible with the MRC/MRT scheme.}
%converge to the same outage probability, but with different convergence rates.
\end{abstract}

%
%
%\begin{minipage}{1\columnwidth}
%\vspace{0.3cm}
%{\scriptsize Corresponding author: Caijun Zhong, Email: caijunzhong@zju.edu.cn\\
% G. Zhu, C. Zhong, and Z. Zhang are with the Institute of Information and Communication Engineering, Zhejiang University, Hangzhou, China\\
%H. A. Suraweera is with the University of Peradeniya, Peradeniya 20400, Sri Lanka. Email: himal@ee.pdn.ac.lk\\
%C. Yuen is with the Singapore University of Technology and Design, 20 Dover
%Drive, Singapore 138682. Email: yuenchau@sutd.edu.sg}
%\end{minipage}

\section{Introduction}\label{section:1}
The relay channel was first introduced by Van der Meulen in 1971 \cite{V.der}. Later, in the seminal work of \cite{T.Cover}, Cover and El Gamal laid foundations to the information-theoretic understanding of the relay channel. The attention on relay channels was recently rekindled as a means to improve the coverage and link reliability in the context of cooperative wireless communications systems \cite{J.Laneman}. Various relaying methods have been proposed in the literature\cite{J.Laneman}, among which,
the amplify-and-forward (AF) protocol is the most popular one, due to its simplicity and low-cost implementation. In AF systems, the relay simply forwards a scaled version of the received noisy signal
from the source to the destination.

To improve the spectrum efficiency, future cellular systems are likely to adopt a more aggressive frequency reuse strategy, which will inevitably result in an interference-limited communication environment \cite{J.H.Winter}. When the relay technology is adopted in cellular systems \cite{K.Loa}, the interference environment becomes increasingly complex. Motivated by the need to understand the performance limitations, a number of works investigating the impact of co-channel interference (CCI) on the performance of relay systems have appeared. For example, \cite{I.Krikidis,I.Krikidis2} studied the performance of relay selection for AF systems with CCI. Assuming Rayleigh fading channels, \cite{C.Zhong} examined the outage probability of dual-hop fixed-gain AF relaying systems with an interference-limited destination, and
\cite{H.Suraweera} studied the outage probability and the average bit error rate of dual-hop variable-gain AF relaying systems with an interference-limited
relay. Later, a scenario considering the more general Nakagami-$m$ fading model was investigated in \cite{Fawaz,Himal}. Moreover, different cases with CCI at both the relay and the destination nodes have been investigated in \cite{W.Xu,S.Ikki0,AM}. More recent works have also investigated the effect of CCI on single antenna
two-way relaying systems for the decode-and-forward protocol \cite{X.Liang} and the AF protocol \cite{M.Matthaiou}. However, it is worth noting that all these prior works deal with the case where all nodes are equipped with a single antenna.

It is well known that the multiple-input multiple-output (MIMO) technology provides extra spatial degrees of freedom which can be efficiently utilized for interference cancellation. To this end, MIMO has been identified as one of the key enabling physical layer technologies in wireless standards such as LTE-Advanced and IMT-Advanced \cite{Q.Limag}. Despite the importance, so far only a few papers have investigated the impact of CCI in MIMO relaying systems \cite{C.Zhong1,K.Hemachandra,H.Ding}.

%Assuming a single interferer at the relay, \cite{C.Zhong1} studied the setup where only one of the source, relay or destination nodes is equipped with multiple antennas, and derived exact analytical expressions and simple high signal-to-noise ratio (SNR) approximations for the outage probability. Meanwhile, \cite{H.Ding} considered the scenario with multi-antenna source and destination and single antenna relay, and analyzed the outage performance of the system assuming interference-limited relay and destination.

In this paper, we consider the scenario with multiple antennas at the relay, and single antenna at the source and destination. This particular system setup studied in the relay communication literature \cite{C.Zhong1,X.Tang} is applicable in several practical scenarios where two nodes (e.g., machine-to-machine type low cost devices) exchange information with the
assistance of an advanced terminal such as a cellular
base-station/cluster-head sensor. {At this point, it is important to highlight the major differences between the current paper and state-of-the art in the literature. Unlike \cite{C.Zhong1} which only considered a single interferer, the current paper allows for arbitrary number of interferers at the relay node. Compared with \cite{H.Ding}, which assumed an interference-limited single antenna relay, the current work considers a multiple antenna relay. Compared with \cite{K.Hemachandra}, which investigated the outage performance of the scenario with interference-limited multiple antenna relay, the current work considers the more general setup by taking into account of the effect of additive white Gaussian noise (AWGN) at the relay. More importantly, in contrast to \cite{C.Zhong1,K.Hemachandra,H.Ding}, where the simple maximum ratio transmission (MRT) and maximum ratio combining (MRC) schemes were used, the current paper adopts more sophisticated linear combining schemes to suppress the CCI.}

In the presence of CCI, linear diversity combining schemes have been widely adopted in the multiple antenna systems because of the low complexity and good performance \cite{A.Shah1}. In the same spirit, in this paper we propose a heuristic two-stage relay processing scheme, i.e., the relay first applies linear combining methods to suppress the CCI, and then forwards the transformed signal to the destination by using the MRT scheme. Three popular linear combining methods, i.e., MRC, zero-forcing (ZF) and minimum mean square error (MMSE), are investigated. To the best of the authors' knowledge, the analysis of diversity combining schemes for the suppression of CCI in dual-hop AF relaying systems has not been presented in the existing literature.

{We present a detailed performance analysis of the considered MRC/MRT, ZF/MRT and MMSE/MRT schemes in Rayleigh fading channels. Our main contributions are summarized as follows:}
\begin{itemize}
\item For the MRC/MRT scheme, we derive a new exact expression involving a single integral for the outage probability of the system, and present a tight closed-form outage lower bound. In addition, we obtain a simple high signal-to-noise ratio (SNR) outage approximation, and prove that the MRC/MRT scheme achieves a diversity order of $N$, where $N$ is the number of antennas at the relay.
\item For the ZF/MRT scheme, we first obtain the optimal combining vector maximizing the end-to-end signal-to-interference-and-noise ratio (SINR) subject to the ZF constraint, and then derive a new exact closed-form expression for the outage probability. We also characterize the high SNR outage behavior and show that it achieves a diversity order of $N-M$, where $M$ is the number of interferers.
\item For the MMSE/MRT scheme, we derive a new exact expression involving a single integral for the outage probability, and propose a tight closed-form outage lower bound. We also characterize the high SNR outage behavior of the MMSE/MRT scheme, and show that it achieves a diversity order of $N$.
\item Our results suggest that the MMSE/MRT scheme always attains the best outage performance, and the ZF/MRT scheme outperforms the MRC/MRT scheme in the low SNR regime, while the MRC/MRT scheme achieves a superior outage performance than the ZF/MRC scheme in the high SNR regime.
\item {We also look into the large $N$ regime,\footnote{The large $N$ regime analysis is of great interest due to the advent of large-MIMO (massive MIMO) technology \cite{Marzetta,Himalrelay}.} and demonstrate that in this case, both the ZF/MRT and MMSE/MRT schemes are capable of completely eliminating the CCI, while perfect interference cancelation is not possible with the MRC/MRT scheme.}
   % all the three schemes achieve the same asymptotic outage performance. However, the rates of different schemes converging to the limit differ significantly, i.e., the MMSE/MRT and ZF/MRT schemes approach the limit with a similar rate, which is faster than that of the MRC/MRT scheme.
\end{itemize}

The remainder of the paper is organized as follows: Section II introduces the system model. Section III presents a detailed investigation of the outage probability achieved by the three different schemes. Numerical results
and discussions are provided in Section IV. Finally, Section V concludes the paper and summarizes the key findings.

{\it Notation}: We use bold upper case letters to denote matrices, bold lower case letters to denote vectors and lower case
letters to denote scalars. {${\left\| {\bf{h}} \right\|_F}$ denotes the Frobenius norm}, ${\tt E}\{x\}$ stands for the expectation of the random variable $x$, ${*}$ denotes the conjugate operator, while $T$ denotes the transpose operator and ${\dag}$ denotes the conjugate transpose operator. ${{\cal CN} (0,1)}$ denotes a scalar complex circular Gaussian random variable with zero mean and unit variance. ${{\bf{I}}_k}$ is the identity matrix of size $k$. $\Gamma(x)$ is the gamma function and $K_v(x)$ is the $v$-th order modified Bessel function of the second kind \cite[Eq. (8.407.1)]{Tables}. $\Gamma \left( {\alpha ,x} \right)$ is the upper incomplete gamma function \cite[Eq. (8.350.2)]{Tables} and ${}_2{F_1}(a,b;c;z)$ is the Gauss Hypergeometric Function \cite[Eq. (9.100)]{Tables}.

\section{System Model}
\begin{figure}[ht]
\centering
\includegraphics[scale=0.7]{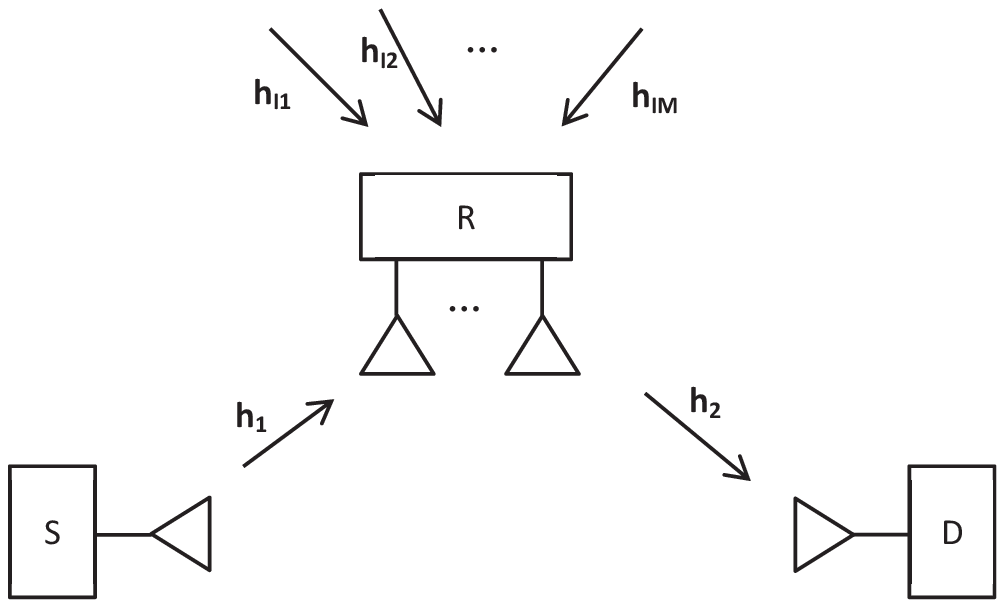}
\caption{System model: S, R and D denote the source, the relay and the destination, respectively.}\label{fig:fig1}
\end{figure}

Let us consider a dual-hop multiple antenna AF relaying system as illustrated in Fig. \ref{fig:fig1}, where both the source and destination are equipped with a single antenna, while the relay is equipped with $N$ antennas. We consider the scenario where the relay is subjected to $M$ independently but not necessarily identically distributed co-channel interferers and AWGN, while the destination is corrupted by AWGN only\footnote{{This scenario is also particularly relevant to frequency-division relay
systems \cite{R.Pabst} where the relay and the destination experience different interference patterns.}}. We also assume that the direct link between the source and the destination does not exist due to obstacles or path loss attenuation/severe fading.

In this paper, we consider half-duplex relaying, and hence a complete transmission occurs in two phases. During the first phase, the source transmits the signal to the relay, and the signal received at the relay is given by
\begin{align}\label{SM:1}
{{{\bf y}}_{{r}}} = {{{\bf h}}_{{1}}}x + \sum\limits_{i = 1}^M {{{{\bf h}}_{{Ii}}}{s_{Ii}} + {{{\bf n}}_{{1}}}} ,
\end{align}
where the ${N\times 1}$ vector ${{{\bf h}}_{{1}}}$ denotes the channel for the source-relay link. The entries of ${{{\bf h}}_{{1}}}$ follow identically and independently distributed (i.i.d.) ${{\cal CN} (0,1)}$. The ${N\times 1}$ vector ${{{\bf h}}_{{Ii}}}$ denotes the channel for the \emph{i}-th interference-relay link, and its entries follow  i.i.d. ${{\cal CN} (0,1)}$, and $x$ is the source symbol satisfying ${\tt E}\left\{ {x{x^* }} \right\} = P_s$. The \emph{i}-th interference symbol is ${s_{Ii}}$ with ${\tt E}\left\{ {{s_{Ii}}s_{Ii}^*} \right\} = {P_{Ii}}$, ${{{\bf n}}_{{1}}}$ is an ${N\times 1}$ vector and denotes the AWGN at the relay node with ${\tt E}\{\mathbf{n}_1\mathbf{n}_1^{\dag}\} = N_0{\mathbf{I}}$.

In the second phase, the relay node transmits a transformed version of the received signal to the destination, and the signal at the destination can be expressed as
\begin{align}\label{SM:2}
y_d = {\bf{h}}_{{2}}{\bf{W}}{{\bf{y}}_{{r}}} +n_2 ,
\end{align}
where ${{\bf{h}}_{{2}}}$ is a ${1\times N}$ vector and denotes the channel for the relay-destination link, and its entries follow i.i.d. ${{\cal CN} (0,1)}$, ${n_2}$ is the AWGN with ${\tt E}\{n_2^{*}n_2\} = N_0$, $\mathbf{W}$ is the transformation matrix at relay node with ${\tt E}\{\left\|\mathbf{W}\mathbf{y}_r\right\|_F^2\} = P_r$.

Combining (\ref{SM:1}) and (\ref{SM:2}), the end-to-end SINR of the system can be computed as
\begin{align}\label{SINR:1}
\gamma  = \frac{{{\left| {{\bf{h}}_{2}{\bf{W}}{{\bf{h}}_{1}}} \right|}^2}P_s}{{\sum\limits_{i = 1}^M {{\left| {{{\bf h}}_{2}{\bf{W}}{{{\bf h}}_{Ii}}} \right|}^2}{P_{Ii}} + {\left\| {{{\bf h}}_{{2}}{\bf{W}}} \right\|}_F^2}{N_0} + {N_0}}.
\end{align}
The optimal relay precoder matrix ${\bf W}$ maximizing the end-to-end SINR $\gamma$ does not seem to be analytically tractable, due to the non-convex nature of the problem. Hence, in this paper, we propose a heuristic two-stage relay processing strategy, i.e., the relay first performs some linear combining method to suppress the CCI, and then forwards the transformed signal to the destination using the MRT scheme since it maximizes the SNR of the relay-destination link. Therefore, the heuristic relay precoder ${\bf W}$ is a rank-1 matrix, i.e., ${\bf{W}} = \omega \frac{{{\bf{h}}_2^{\bf{\dag }}}}{{\left\| {{{\bf{h}}_2}} \right\|}_F}{{\bf{w}}_1}$, where $\omega$ is the power constraint factor, $\frac{{{\bf{h}}_2^{\bf{\dag }}}}{{\left\| {{{\bf{h}}_2}} \right\|}_F}$ is used for matching the second hop channel and ${{\bf{w}}_1}$ is a ${1\times N}$ linear combining vector, which depends on the linear combining scheme employed by the relay and will be specified in the following section.

\section{Outage Probability Analysis}
In this section, we investigate the outage probability of the MRC/MRT, ZF/MRT and MMSE/MRT schemes. New exact analytical expressions are derived for the outage probability of all three schemes. In addition, simple high SINR approximations are presented, which provide a concise characterization of the achievable diversity order of the system, and enable a performance comparison of the three schemes.

The outage probability is an important performance metric, which is defined as the instantaneous SINR falls below a pre-defined threshold $\gamma_{\sf th}$. Mathematically, it can be expressed as
\begin{align}
{P_{\sf out}} = {\mathop{\rm Prob}\nolimits} \left( {\gamma  < {\gamma _{\sf th}}} \right).
\end{align}

\subsection{MRC/MRT Scheme}
For the MRC/MRT scheme, ${{\bf{w}}_1}$ is set to match the first hop channel, hence, ${{\bf{w}}_1}=\frac{{{\bf{h}}_1^{\bf{\dag }}}}{{\left\| {{{\bf{h}}_1}} \right\|}_F}$.
%
%Therefore, it can be easily noted that this scheme does not require the availability of the interference channel information (ICI) at the relay node and is convenient to be used by the relays with limited resources.
To meet the transmit power constraint at the relay, the constant $\omega^2$ can be computed as
\begin{align}
{\omega ^2} = \frac{{{P_r}}}{{{{\bf h}}_{{1}}^{\bf{\dag }}{{{\bf h}}_{{1}}}P_s + \frac{{\sum\limits_{i = 1}^M {{{\left| {{{\bf h}}_{{1}}^{{\dag }}{{{\bf h}}_{{{Ii}}}}} \right|}^2}{P_{Ii}}} }}{{{{\left\| {{{{\bf h}}_{{1}}}} \right\|}_F^2}}} + {N_0}}}.
\end{align}
Therefore, the end-to-end SINR for the MRC/MRT scheme, $\gamma_{\sf MRC}$ can be expressed as (\ref{PA:1}).
\begin{figure*}
\begin{align}\label{PA:1}
\gamma_{\sf MRC}  = \frac{{{{\left| {{{\bf{h}}_{\bf{2}}}} \right|}^2}{{\left| {{{\bf{h}}_{\bf{1}}}} \right|}^2}P_s}}{{{{\left| {{{\bf{h}}_{\bf{2}}}} \right|}^2}\frac{{\sum\limits_{i = 1}^M {{{\left| {{\bf{h}}_{\bf{1}}^{\bf{\dag }}{{\bf{h}}_{{\bf{Ii}}}}} \right|}^2}{P_{Ii}}} }}{{{{\left| {{{\bf{h}}_{\bf{1}}}} \right|}^2}}} + {{\left| {{\bf{h}}_{\bf{2}}^{\bf{\dag }}} \right|}^2}{N_0} + \frac{{{N_0}}}{{{P_r}}}\left( {{{\left| {{{\bf{h}}_{\bf{1}}}} \right|}^2}P_s + \frac{{\sum\limits_{i = 1}^M {{{\left| {{\bf{h}}_{\bf{1}}^{\bf{\dag }}{{\bf{h}}_{{\bf{Ii}}}}} \right|}^2}{P_{Ii}}} }}{{{{\left| {{{\bf{h}}_{\bf{1}}}} \right|}^2}}} + {N_0}} \right)}}.
\end{align}
\hrule
\end{figure*}
Now, with the end-to-end SINR given in (\ref{PA:1}), we are ready to establish the outage probability of the MRC/MRT scheme.
For notational convenience, we define ${\rho _1} = \frac{{P_s}}{{{N_0}}}$, ${\rho _2} = \frac{{{P_r}}}{{{N_0}}}$ and ${\rho _{Ii}} = \frac{{{P_{Ii}}}}{{{N_0}}}$, $i=1,\dots,M$. We have the following key result.
\begin{theorem}\label{theorem:1}
In the presence of interferers at the relay, the outage probability of the dual-hop AF relaying system with the MRC/MRT scheme can be expressed as
\begin{multline}
{P_{\sf out}^{\sf MRC}} = 1 - \frac{{2{e^{ - \frac{{{\gamma _{\sf th}}}}{{{\rho _1}}} - \frac{{{\gamma _{\sf th}}}}{{{\rho _2}}}}}}}{{\Gamma \left( N \right)}}\sum\limits_{m = 0}^{N - 1} {{\left( {\frac{{{\gamma _{\sf th}}}}{{{\rho _1}}}} \right)}^m}\frac{1}{{m!}}\\
\sum\limits_{j = 0}^m {{m\choose j}{{\left( {\frac{1}{{{\rho _2}}}} \right)}^{m - j}}\sum\limits_{k = 0}^{N + j - 1} {N+j-1\choose k} }  \\
\times{\left( {\frac{{{\gamma _{\sf th}}}}{{{\rho _2}}}} \right)^{N + j - 1 - k}}{\left[ {\frac{{\left( {{\gamma _{\sf th}} + 1} \right){\gamma _{\sf th}}}}{{{\rho _1}{\rho _2}}}} \right]^{\frac{{k - m + 1}}{2}}}{\cal I}\left( {{\gamma _{\sf th}}} \right),
\end{multline}
with ${\cal I}\left( {{\gamma _{\sf th}}} \right)$ given in (\ref{multline:5}),
\begin{figure*}
\begin{multline}\label{multline:5}
{\cal I}\left( {{\gamma _{\sf th}}} \right) = \sum\limits_{p = 1}^{\rho ({\bf{D}})} {\sum\limits_{q = 1}^{{\tau _p}({\bf{D}})} {{\chi _{p,q}}({\bf{D}})\frac{{\rho _{I\left\langle p \right\rangle }^{ - q}}}{{(q - 1)!}}\int_0^\infty  {{K_{k - m + 1}}\left( {2\sqrt {\frac{{\left( {{\gamma _{\sf th}} + 1} \right){\gamma _{\sf th}}}}{{{\rho _1}{\rho _2}}}\left( {x + 1} \right)} } \right)} } }{\left( {x + 1} \right)^{\frac{{k + m + 1}}{2}}}{x^{q - 1}}{e^{ - \left( {\frac{{{\gamma _{\sf th}}}}{{{\rho _1}{N_0}}} + \frac{1}{{{\rho _{I\left\langle p \right\rangle }}}}} \right)x}}dx.
\end{multline}
\hrule
\end{figure*}
where $\mathbf{D} = {\sf diag}({\rho _{I1}},{\rho _{I2}}, \cdots ,{\rho _{IM}})$, $\rho (\mathbf{D})$ is the number of distinct diagonal elements of $\mathbf{D}$, ${\rho _{I\left\langle 1 \right\rangle }} > {\rho _{I\left\langle 2 \right\rangle }} >  \cdots  > {\rho _{I\left\langle {\rho (\mathbf{D})} \right\rangle }}$ are the distinct diagonal elements in decreasing order, ${\tau _i}(\mathbf{D})$ is the multiplicity of ${\rho _{I\left\langle i \right\rangle }}$ and ${\chi _{i,j}}(\mathbf{D})$ is the ${\left( {i,j} \right)-\mbox{th}}$ characteristic coefficient of $\mathbf{D}$.

\proof See Appendix \ref{appendix:theorem:1}. \endproof
\end{theorem}
%
%
%For the case where  $\{\rho _{Ii}\}$ are all equal, i.e., ${\rho _{Ii}} \buildrel \Delta \over = {\rho _I}$ for any \emph{i}, (\ref{multline:5}) reduces to
%\begin{multline}
%{\cal I}\left( {{\gamma _{th}}} \right) = \frac{{\rho _{I}^{ - M}}}{{(M - 1)!}}\int_0^\infty  {{K_{k - m + 1}}\left( {2\sqrt {\frac{{\left( {{\gamma _{th}} + 1} \right){\gamma _{th}}}}{{{\rho _1}{\rho _2}}}\left( {\frac{{{x}}}{{{N_0}}} + 1} \right)} } \right){{\left( {\frac{{{x}}}{{{N_0}}} + 1} \right)}^{\frac{{k + m + 1}}{2}}}} \\
%\times x^{M - 1}{e^{ - \left( {\frac{{{\gamma _{th}}}}{{{\rho _1}{N_0}}} + \frac{1}{{{\rho _{I}}}}} \right){x}}}d{x}.
%\end{multline}

Theorem \ref{theorem:1} presents the exact outage probability of the MRC/MRT scheme, which is quite general and valid for the system with arbitrary number of antennas and interferers. For the special case with a single interferer, Theorem \ref{theorem:1} reduces to the result derived in \cite[Theorem 13]{C.Zhong1}.
To the best of the authors' knowledge, the integral ${\cal I}$ does not admit a closed-form expression. However, this
single integral expression can be efficiently evaluated numerically using software such as Matlab or MATHEMATICA, which still provides computational advantage over
a Monte Carlo simulation method.

Alternatively, we can use the following closed-form lower bound of the outage
probability, which is tight across the entire SNR range, and becomes exact in the high SNR regime.

\begin{corollary}\label{coro:1}
In the presence of interferers at the relay, the outage probability of the dual-hop AF relaying system with the MRC/MRT scheme is lower bounded by
\begin{multline}\label{align:c1}
P_{\sf out}^{\sf lMRC} = 1 - \frac{{\Gamma \left( {N,\frac{{{\gamma _{\sf th}}}}{{{\rho _2}}}} \right)}}{{\Gamma \left( N \right)}}{e^{ - \frac{{{\gamma _{\sf th}}}}{{{\rho _1}}}}}\sum\limits_{k = 0}^{N - 1} {{\left( {\frac{{{\gamma _{\sf th}}}}{{{\rho _1}}}} \right)}^k}\frac{1}{{k!}}\sum\limits_{l = 0}^k {k\choose l}
\\\sum\limits_{i = 1}^{\rho (\mathbf{D})} {\sum\limits_{j = 1}^{{\tau _i}(\mathbf{D} )} {{\chi _{i,j}}(\mathbf{D} )\frac{{\Gamma \left( {j + l} \right)}}{{\Gamma \left( j \right)}}}}\rho _{I\langle i\rangle}^l{{\left( {\frac{{{\rho _1}}}{{{\rho _1} + {\rho _{I\langle i\rangle}}{\gamma _{\sf th}}}}} \right)}^{j + l}}.
\end{multline}
\end{corollary}
\proof See Appendix \ref{appendix:corollary:1}. \endproof

%Eq. (\ref{align:c1}) admits simpler forms under special cases. For example, in the case where all interferers are clustered together with $\{\rho _{Ii}\}$ are all equal, i.e., ${\rho _{Ii}} \buildrel \Delta \over = {\rho _I}$ for any \emph{i}, (\ref{align:c1}) reduces to
%\begin{align}
%P_{out}^{low}\left( {{\gamma _{th}}} \right) = 1 - \frac{{\Gamma \left( {N,\frac{{{\gamma _{th}}}}{{{\rho _2}}}} \right)}}{{\Gamma \left( N \right)}}{e^{ - \frac{{{\gamma _{th}}}}{{{\rho _1}}}}}\sum\limits_{k = 0}^{N - 1} {{{\left( {\frac{{{\gamma _{th}}}}{{{\rho _1}}}} \right)}^k}\frac{1}{{k!}}\sum\limits_{l = 0}^k {\left( {\begin{array}{*{20}{c}}
%k\\
%l
%\end{array}} \right)\frac{{\Gamma \left( {M + l} \right)}}{{\Gamma \left( M \right)}}\rho _I^l{{\left( {\frac{{{\rho _1}}}{{{\rho _1} + {\rho _I}{\gamma _{th}}}}} \right)}^{M + l}}} }.
%\end{align}
While Theorem \ref{theorem:1} and Corollary \ref{coro:1} provide efficient methods for evaluating the exact outage probability of the system, these expressions are quite complicated, and do not allow for easy extraction of useful insights. Motivated by this, we now look into the high SNR regime, and derive a simple approximation for the outage probability, which enables the characterization of the achievable diversity order of the MRC/MRT scheme.
\begin{theorem}\label{theorem:2}
In the high SNR regime, i.e., ${\rho _2} = \mu {\rho _1}$, ${\rho _1} \to \infty $, {with $\mu$ being a finite constant}, the outage probability of dual-hop AF relaying system with the MRC/MRT scheme can be approximated as
\begin{align}\label{align:2}
&P_{\sf out}^{\sf MRC} \approx \notag\\
&\left[ {\frac{1}{{{\mu ^N}}} + \sum\limits_{k = 0}^N {N\choose k} \sum\limits_{i = 1}^{\rho (\mathbf{D})} {\sum\limits_{j = 1}^{{\tau _i}(\mathbf{D} )} {{\chi _{i,j}}(\mathbf{D} )\frac{{\Gamma \left( {k + j} \right)}}{{\Gamma \left( j \right)}}\rho _{I\langle i\rangle}^k} } } \right]\notag\\
&\times\frac{{\left( {\frac{{{\gamma _{\sf th}}}}{{{\rho _1}}}} \right)^N}}{{\Gamma \left( {N + 1} \right)}} + { o}\left({\left( {\frac{{{\gamma _{\sf th}}}}{{{\rho _1}}}} \right)^{N+1}}\right).
\end{align}
\proof See Appendix \ref{appendix:theorem:2}. \endproof
\end{theorem}
For the special case where  $\{\rho _{Ii}\}$ are equal, i.e., ${\rho _{Ii}} \buildrel \Delta \over = {\rho _I}$ for any \emph{i}, (\ref{align:2}) reduces to
\begin{align}\label{MRC:1}
&P_{\sf out}^{\sf MRC} = \frac{1}{{\Gamma \left( {N + 1} \right)}}\left[ {\frac{1}{{{\mu ^N}}} + \sum\limits_{k = 0}^N {N\choose k} \frac{{\Gamma \left( {k + M} \right)}}{{\Gamma \left( M \right)}}\rho _I^k} \right]\notag\\
&\times{\left( {\frac{{{\gamma _{\sf th}}}}{{{\rho _1}}}} \right)^N}+
  { o}\left({\left( {\frac{{{\gamma _{\sf th}}}}{{{\rho _1}}}} \right)^{N+1}}\right).
\end{align}

Theorem \ref{theorem:2} indicates that the  MRC/MRT scheme achieves a diversity order of $N$. Moreover, it implies that the number of interferers does not affect the achievable diversity order, it however, causes a detrimental effect on the array gain. This key observation suggests that, in the presence of strong CCI, the outage performance of the MRC/MRT scheme will be significantly affected. Hence, in such a scenario, the MRC/MRT scheme may not be suitable. {Motivated by this, we now study the performance of more sophisticated linear combining techniques with superior interference suppression capability, namely, the ZF/MRT scheme and the MMSE/MRT scheme.}

\subsection{ZF/MRT Scheme}
In the ZF/MRT scheme, the relay utilizes the available multiple antennas to completely eliminate the CCI.\footnote{We would like to point out that the performance of ZF scheme in multiple antenna dual-hop AF systems has been studied in \cite{R.Louie}, where ZF is applied for inter-stream interference cancellation. To the best of the authors' knowledge, application of ZF for CCI cancellation in dual-hop AF relaying systems has not been studied.} To ensure this is possible, the number of the antennas equipped at the relay should be greater than the number of interferers. Hence, for the ZF/MRT scheme, it is assumed that $N > M$.

Define an $N\times M$ matrix ${{\bf{H}}_I} = \left[ {{{\bf{h}}_{I1}},{{\bf{h}}_{I2}} \cdots {{\bf{h}}_{IM}}} \right]$ as the interference channel matrix, the SINR expression in (\ref{SINR:1}) can be alternatively expressed as
\begin{align}\label{SINR:2}
 \gamma_{\sf ZF}= \frac{{{\omega ^2}{{\left\| {{{\bf{h}}_2}} \right\|}_F^2}{{\left| {{{\bf{w}}_1}{{\bf{h}}_1}} \right|}^2}{\rho_1}}}{{{\omega ^2}{{\left\| {{{\bf{h}}_2}} \right\|}_F^2}{{\left( {{\bf{w}}_1}{{\bf{H}}_I}{{\bf{D}}}{{\bf{H}}_I^{\dag}}{\bf w}_1^{\dag} \right)}} + {\omega ^2}{{\left\| {{{\bf{h}}_2}} \right\|}_F^2}{{\left\| {{{\bf{w}}_1}} \right\|}_F^2} + 1}}.
\end{align}
Hence, the optimal combining vector ${{\bf{w}}_1}$ should be the solution of the following maximization problem
\begin{align}\label{ZF:1}
\begin{array}{l}
{{\bf{w}}_1} = \arg \mathop {\max }\limits_{{\bf{w}}_1} \;\gamma_{\sf ZF}\\
{\text{s.t.}}\;\;{{\bf{w}}_1}{{\bf{H}}_I} = {\bf{0}}\;\;\& \;\;\left\| {{\bf{w}}_1} \right\|_F = 1.
\end{array}
\end{align}
The problem in (\ref{ZF:1}) can be solved as follows:
\begin{proposition}\label{prop:11}
The optimal combining vector ${\bf w}_1$ is given by
 \begin{align}
 {\bf{w}}_1 = \frac{{{\bf{h}}_1^\dag {\bf{P}}}}{{\sqrt {{\bf{h}}_1^\dag {\bf{P}}{{\bf{h}}_1}} }},
 \end{align}
 where ${\bf{P}} = {{\bf{I}}_N} - {{\bf{H}}_I}{\left( {{\bf{H}}_I^\dag {{\bf{H}}_I}} \right)^{ - 1}}{\bf{H}}_I^\dag $.
\end{proposition}

\proof See Appendix \ref{app:prop:11}. \endproof

%Utilizing the SNR expression in , the original objective function in (\ref{ZF:1}) can be expressed as
%\begin{align}\label{ZF:2}
%\begin{array}{l}
%{{\bf{w}}_1} = \arg \;\mathop {\max }\limits_{{{{\bf{w}}_1}}} \;\;\frac{{{\omega ^2}{{\left| {{{\bf{h}}_2}} \right|}^2}{{\left| {\;{{{\bf{w}}_1}}{{\bf{h}}_1}} \right|}^2}P}}{{{\omega ^2}{{\left| {{{\bf{h}}_2}} \right|}^2}{N_0} + {N_0}}}\\
%{\text{s.t.}}\;\;{{{\bf{w}}_1}}{{\bf{H}}_I} = {\bf{0}}\;\;\& \;\;\left| {{{{\bf{w}}_1}}} \right| = 1.
%\end{array}
%\end{align}

%It is easy to notice that the maximization problem in (\ref{ZF:2}) can be further simplified as
%
%Now, starting from (\ref{ZF:3}) we can finally obtain the solution for this maximization problem as follows.

%Hence, in order to meet the power constraint at the relay node, we have
%

Having obtained the optimal combining vector $\mathbf{w}_1$, the end-to-end SINR can be expressed as
\begin{align}\label{ZF:5}
\gamma_{\sf ZF} = \frac{{{{\left\| {{{\bf{h}}_2}} \right\|}_F^2}\left| {{\bf{h}}_1^{\bf{\dag }}{\bf{P}}{{\bf{h}}_1}} \right|{\rho_1}{\rho_2}}}{{{{\left\| {{{\bf{h}}_2}} \right\|}_F^2}{\rho_2} + \left( {\left| {{\bf{h}}_1^{\bf{\dag }}{\bf{P}}{{\bf{h}}_1}} \right|{\rho_1} + 1} \right)}}.
\end{align}
With the above SINR expression, we now study the outage probability of the ZF/MRT scheme.
\begin{theorem}\label{theorem:3}
In the presence of interferers at the relay, the outage probability of the dual-hop AF relaying system with the ZF/MRT scheme can be expressed as
\begin{multline}\label{ZF:4}
P _{\sf out}^{\sf ZF} = 1 - \frac{{2{e^{ - \frac{{{\gamma _{\sf th}}}}{{{\rho _1}}} - \frac{{{\gamma _{\sf th}}}}{{{\rho _2}}}}}}}{{\Gamma \left( N \right)}}{\sum\limits_{m = 0}^{N - M - 1} {\left( {\frac{{{\gamma _{\sf th}}}}{{{\rho _1}}}} \right)} ^m}\frac{1}{{m!}}\sum\limits_{j = 0}^m {m\choose j}\\
 \times{\left( {\frac{1}{{{\rho _2}}}} \right)^{m - j}}\sum\limits_{k = 0}^{N + j - 1} {N+j-1\choose k} {\left( {\frac{{{\gamma _{\sf th}}}}{{{\rho _2}}}} \right)^{N + j - k - 1}}\\
 \times{\left( {\frac{{\left( {1 + {\gamma _{\sf th}}} \right){\gamma _{\sf th}}}}{{{\rho _1}{\rho _2}}}} \right)^{\frac{{k - m + 1}}{2}}}{{K} _{k - m + 1}}\left( {2\sqrt {\frac{{\left( {1 + {\gamma _{\sf th}}} \right){\gamma _{\sf th}}}}{{{\rho _1}{\rho _2}}}} } \right).
\end{multline}
\proof
{We start by expressing the end-to-end SINR given in (\ref{ZF:5}) as
\begin{align}
\gamma_{\sf ZF} = \frac{{{y_2}{y_3}{\rho _1}{\rho _2}}}{{{y_2}{\rho _2} + {y_3}{\rho _1} + 1}},
\end{align}
where $y_2={{\left\| {{{\bf{h}}_2}} \right\|}_F^2}$ and ${y_3} = {\left| {{\bf{h}}_1^\dag {\bf{P}}{{\bf{h}}_1}} \right|}$. From \cite{Z.Ding}, the probability density function (p.d.f.) of $y_3$ can be expressed as
\begin{align}
{f_{{y_3}}}\left( x \right) = \frac{{{x^{N - M - 1}}}}{{\left( {N - M - 1} \right)!}}{e^{ - x}}.
\end{align}
Hence, the outage probability of the ZF/MRT scheme can be written as
\begin{align}
{P_{\sf out}^{\sf ZF}}& = {\mathop{\rm Prob}\nolimits}\left( {\frac{{{y_2}{y_3}{\rho _1}{\rho _2}}}{{{y_2}{\rho _2} + {y_3}{\rho _1} + 1}} \le {\gamma _{\sf th}}} \right) \notag\\
&= {\mathop{\rm Prob}\nolimits}\left( {{y_3}\frac{{{y_2} - \frac{{{\gamma _{\sf th}}}}{{{\rho _2}}}}}{{{y_2} + \frac{1}{{{\rho _2}}}}} \le \frac{{{\gamma _{\sf th}}}}{{{\rho _1}}}} \right).
\end{align}
To this end, invoking the result of \cite[Lemma 3]{C.Zhong1} yields the desired result.}
\endproof
\end{theorem}

Theorem \ref{theorem:3} provides an exact closed-form expression for the outage probability of the ZF/MRT scheme. This expression only involves standard mathematical functions and hence, can be efficiently evaluated. To gain further insights, we now look into the high SNR regime, and present a simple and informative approximation for the outage probability.

\begin{theorem}\label{theorem:4}
In the high SNR regime, i.e., ${\rho _2} = \mu {\rho _1}$, ${\rho _1} \to \infty $, the outage probability of the dual-hop AF relaying system with the ZF/MRT scheme can be approximated as
\begin{align}\label{ZF:6}
{P_{\sf out}^{\sf ZF}} = \frac{1}{{\Gamma \left( {N - M + 1} \right)}}{\left( {\frac{{{\gamma _{\sf th}}}}{{{\rho _1}}}} \right)^{N - M}} + { o}\left({\left( {\frac{{{\gamma _{\sf th}}}}{{{\rho _1}}}} \right)^{N-M+1}}\right).
\end{align}
\proof See Appendix \ref{appendix:theorem:4}. \endproof
\end{theorem}

As expected, we see that the interference power does not affect the outage probability of the ZF/MRT scheme. It is also interesting to observe that $\mu$ does not affect the outage probability at high SNR. In addition, Theorem \ref{theorem:4} indicates that the achievable diversity order of the ZF/MRT scheme is $N-M$. Compared with the MRC/MRT scheme, which attains a diversity order of $N$, the ZF/MRT scheme incurs a diversity loss of $M$. This important observation suggests that complete elimination of CCI may not be the best option in terms of the outage performance.

\subsection{MMSE/MRT Scheme}
The ZF scheme completely eliminates the CCI at the relay, which however may cause an elevated noise level. In contrast, the MMSE scheme does not fully eliminate the CCI, instead, it provides the optimum trade-off between interference suppression and noise enhancement. In the following, we study the outage performance of the MMSE/MRT scheme.
%is a kind of interference-cancelation technique which further take into account the interference and the noise effect. Therefore, in general, MMSE combining can achieve better performance but requires an estimate of the noise variance at the receiver. Inspired by this, we propose the MMSE/MRT scheme and investigate its performance in this subsection.
{To make the analysis tractable, we assume that ${\rho _{Ii}} \equiv {\rho _I}, \forall i = 1,2, \ldots, M $. It is important to note that the equal interference power assumption adopted to simplify the ensuing analysis is of practical interest as well. For example, it applies when the interference sources are clustered together \cite{D.Costacluster} or when the interference originates from a multiple antenna source implementing an uniform power allocation policy. In addition, we will later illustrate numerically that our analytical results provide very accurate approximations to the outage probability for scenarios with distinct interference power.}

According to the principle of MMSE \cite{H.Gao}, ${\bf w}_1$ is given by
\begin{align}
{{\bf{w}}_1} = {\bf{h}}_1^\dag {\left( {{{\bf{h}}_1}{\bf{h}}_1^\dag  + {{\bf{H}}_I}{\bf{H}}_I^\dag  + \frac{{{1}}}{{{\rho_I}}}{\bf{I}}} \right)^{ - 1}}.
 \end{align}
Also, in order to meet the power constraint at the relay, we have
\begin{align}
{\omega ^2} = \frac{{{\rho_2}}}{{{{\left| {{\bf{w}}_1 {{\bf{h}}_1}} \right|}^2}\rho_1 + \sum\limits_{i = 1}^M {{{\left| {{\bf{w}}_1 {{\bf{h}}_{Ii}}} \right|}^2}} {\rho_I} + {{\left\| {{\bf{w}}_1 } \right\|}_F^2}}}.
\end{align}
Therefore, the end-to-end SINR can be expressed as
%in (\ref{MMSE:1}) shown on the top of the next page.
%\begin{figure*}
%\begin{align}\label{MMSE:1}
%\gamma_{\sf MMSE}  = \frac{{{{\left| {{\bf{w}}_1 {{\bf{h}}_1}} \right|}^2}\rho_1}}{{\left( {{\rm{1 + }}\frac{{1}}{{{\rho_2}{{\left\| {{{\bf{h}}_2}} \right\|}_F^2}}}} \right)\left( {\sum\limits_{i = 1}^M {{{\left| {{\bf{w}}_1 {{\bf{h}}_{Ii}}} \right|}^2}} {\rho_I} + {{\left\| {{\bf{w}}_1 } \right\|}_F^2}} \right) + \frac{{1}}{{{\rho_2}{{\left\| {{{\bf{h}}_2}} \right\|}_F^2}}}{{\left| {{\bf{w}}_1 {{\bf{h}}_1}} \right|}^2}{\rho_1}}}.
%\end{align}
%\hrule
%\end{figure*}
\begin{align}\label{MMSE:1}
&\gamma_{\sf MMSE}  = \notag\\
&\frac{{{{\left| {{\bf{w}}_1 {{\bf{h}}_1}} \right|}^2}\rho_1}}{{\left( {{\rm{1 + }}\frac{{1}}{{{\rho_2}{{\left\| {{{\bf{h}}_2}} \right\|}_F^2}}}} \right)\left( {\sum\limits_{i = 1}^M {{{\left| {{\bf{w}}_1 {{\bf{h}}_{Ii}}} \right|}^2}} {\rho_I} + {{\left\| {{\bf{w}}_1 } \right\|}_F^2}} \right) + \frac{{{\left| {{\bf{w}}_1 {{\bf{h}}_1}} \right|}^2}{\rho_1}}{{{\rho_2}{{\left\| {{{\bf{h}}_2}} \right\|}_F^2}}}}}.
\end{align}

In order to study the outage probability, the remaining task is to characterize the distribution of $\gamma_{\sf MMSE}$. However, the involved SINR expression given in (\ref{MMSE:1}) is difficult to handle. Hence, we first express the SINR as
\begin{align}\label{MMSE:21}
\gamma_{\sf MMSE}  &= \frac{{\rho_2}{{\left\| {{{\bf{h}}_2}} \right\|}_F^2}Z}{{\left({1+\rho_2}{{\left\| {{{\bf{h}}_2}} \right\|}_F^2}\right)+Z}},
\end{align}
where $Z$ is defined as
\begin{align}
Z\buildrel \Delta \over =  \frac{{{{\left| {{\bf{w}}_1 {{\bf{h}}_1}} \right|}^2}{\rho_1}}}{{\left( {\sum\limits_{i = 1}^M {{{\left| {{\bf{w}}_1 {{\bf{h}}_{Ii}}} \right|}^2}} {\rho_I} + {{\left\| {{\bf{w}}_1 } \right\|}_F^2}} \right)}}.
\end{align}
Now, let us focus on $Z$ for the moment. The distribution of the random variable $Z$ has been studied in \cite{H.Gao}, where an exact cumulative distribution function (c.d.f.) expression was presented. However, the final expression is a piecewise function, which is not amenable to further processing, and does not seem to be useful here. To circumvent this difficulty, we first derive an alternative unified expression for the c.d.f. of $Z$.
\begin{proposition}\label{proposition:2}
The c.d.f. of the random variable $Z$ can be expressed as in (\ref{MMSE:5}) shown on the top of the next page,
\begin{figure*}
\begin{align}\label{MMSE:5}
{F _Z}\left( z \right) = 1 - \frac{{\Gamma \left( {N,\frac{z}{{{\rho _1}}}} \right)}}{{\Gamma \left( N \right)}} + \Gamma \left( {M + 1} \right){e^{ - \frac{z}{{{\rho _1}}}}}{\left( {\frac{z}{{{\rho _1}}}} \right)^N}\sum\limits_{m = m_1}^N {\frac{{\rho _I^{N - m + 1}}{}_2{F_1}\left( {M + 1,N - m + 1;N - m + 2; - \frac{{{\rho _I}}}{{{\rho _1}}}z} \right)}{{\Gamma \left( m \right)\Gamma \left( {N - m + 2} \right)\Gamma \left( {m - N + M} \right)}}},
\end{align}
\hrule
\end{figure*}
% \begin{multline}\label{MMSE:5}
%{F _Z}\left( z \right) = 1 - \frac{{\Gamma \left( {N,\frac{z}{{{\rho _1}}}} \right)}}{{\Gamma \left( N \right)}} + \Gamma \left( {M + 1} \right){e^{ - \frac{z}{{{\rho _1}}}}}{\left( {\frac{z}{{{\rho _1}}}} \right)^N}\sum\limits_{m = m_1}^N {\frac{{\rho _I^{N - m + 1}}{}_2{F_1}\left( {M + 1,N - m + 1;N - m + 2; - \frac{{{\rho _I}}}{{{\rho _1}}}z} \right)}{{\Gamma \left( m \right)\Gamma \left( {N - m + 2} \right)\Gamma \left( {m - N + M} \right)}}},
% \end{multline}
where $m_1=\max (0,N - M) + 1$.
\end{proposition}
\proof See Appendix \ref{appendix:proposition:2}. \endproof

Having obtained the c.d.f. expression of $Z$, we are now ready to study the outage probability of the MMSE/MRT scheme, and we have the following key result.
\begin{theorem}\label{theorem:5}
In the presence of interferers at the relay, the outage probability of the dual-hop AF relaying system with the MMSE/MRT scheme can be expressed as in (\ref{MMSE:2}) shown on the top of the next page,
\begin{figure*}
\begin{multline}\label{MMSE:2}
{P_{\sf out}^{\sf MMSE}} = 1 - \frac{{2{e^{ - \frac{{{\gamma _{\sf th}}}}{{{\rho _1}}} - \frac{{{\gamma _{\sf th}}}}{{{\rho _2}}}}}}}{{\Gamma \left( N \right)}}{\sum\limits_{m = 0}^{N - 1} {\left( {\frac{{{\gamma _{\sf th}}}}{{{\rho _1}}}} \right)} ^m}\frac{1}{{m!}}\sum\limits_{j = 0}^m {m\choose j} {\left( {\frac{1}{{{\rho _2}}}} \right)^{m - j}}\sum\limits_{k = 0}^{N + j - 1} {N+j-1\choose k}{\left( {\frac{{{\gamma _{\sf th}}}}{{{\rho _2}}}} \right)^{N + j - k - 1}}\\
\times{\left( {\frac{{\left( {1 + {\gamma _{\sf th}}} \right){\gamma _{\sf th}}}}{{{\rho _1}{\rho _2}}}} \right)^{\frac{{k - m + 1}}{2}}}{K _{k - m + 1}}\left( {2\sqrt {\frac{{\left( {1 + {\gamma _{\sf th}}} \right){\gamma _{\sf th}}}}{{{\rho _1}{\rho _2}}}} } \right) + {e^{ - \frac{{{\gamma _{\sf th}}}}{{{\rho _1}}} - \frac{{{\gamma _{\sf th}}}}{{{\rho _2}}}}}{\left( {\frac{{{\gamma _{\sf th}}}}{{{\rho _1}}}} \right)^N}\frac{{\Gamma \left( {M + 1} \right)}}{{\Gamma \left( N \right)}}\\
\times\sum\limits_{m = m_1}^N {\frac{{\rho _I^{N - m + 1}}}{{\Gamma \left( m \right)\Gamma \left( {N - m + 2} \right)\Gamma \left( {m - N + M} \right)}}}{\sum\limits_{j = 0}^N {{N\choose j}\left( {\frac{1}{{{\rho _2}}}} \right)} ^{N - j}}{\sum\limits_{k = 0}^{N + j - 1} {{N+j-1\choose k}\left( {\frac{{{\gamma _{\sf th}}}}{{{\rho _2}}}} \right)} ^{N + j - 1 - k}}{{\cal {I}}_1}\left( {{\gamma _{\sf th}}} \right),
\end{multline}
\hrule
\end{figure*}
%\begin{multline}\label{MMSE:2}
%{P_{\sf out}^{\sf MMSE}} = 1 - \frac{{2{e^{ - \frac{{{\gamma _{\sf th}}}}{{{\rho _1}}} - \frac{{{\gamma _{\sf th}}}}{{{\rho _2}}}}}}}{{\Gamma \left( N \right)}}{\sum\limits_{m = 0}^{N - 1} {\left( {\frac{{{\gamma _{\sf th}}}}{{{\rho _1}}}} \right)} ^m}\frac{1}{{m!}}\sum\limits_{j = 0}^m {m\choose j} {\left( {\frac{1}{{{\rho _2}}}} \right)^{m - j}}\sum\limits_{k = 0}^{N + j - 1} {N+j-1\choose k}{\left( {\frac{{{\gamma _{\sf th}}}}{{{\rho _2}}}} \right)^{N + j - k - 1}}{\left( {\frac{{\left( {1 + {\gamma _{\sf th}}} \right){\gamma _{\sf th}}}}{{{\rho _1}{\rho _2}}}} \right)^{\frac{{k - m + 1}}{2}}}\\
%\times{K _{k - m + 1}}\left( {2\sqrt {\frac{{\left( {1 + {\gamma _{\sf th}}} \right){\gamma _{\sf th}}}}{{{\rho _1}{\rho _2}}}} } \right) + {e^{ - \frac{{{\gamma _{\sf th}}}}{{{\rho _1}}} - \frac{{{\gamma _{\sf th}}}}{{{\rho _2}}}}}{\left( {\frac{{{\gamma _{\sf th}}}}{{{\rho _1}}}} \right)^N}\frac{{\Gamma \left( {M + 1} \right)}}{{\Gamma \left( N \right)}}\sum\limits_{m = m_1}^N {\frac{{\rho _I^{N - m + 1}}}{{\Gamma \left( m \right)\Gamma \left( {N - m + 2} \right)\Gamma \left( {m - N + M} \right)}}} \\
%\times{\sum\limits_{j = 0}^N {{N\choose j}\left( {\frac{1}{{{\rho _2}}}} \right)} ^{N - j}}{\sum\limits_{k = 0}^{N + j - 1} {{N+j-1\choose k}\left( {\frac{{{\gamma _{\sf th}}}}{{{\rho _2}}}} \right)} ^{N + j - 1 - k}}{{\cal {I}}_1}\left( {{\gamma _{\sf th}}} \right),
%\end{multline}
where ${{\cal {I}}_1}\left( {{\gamma _{\sf th}}} \right)$ can be written as in (\ref{MMSE:6}) also shown on the top of the next page.
\begin{figure*}
\begin{align}\label{MMSE:6}
{{\cal {I}}_1}\left( {{\gamma _{\sf th}}} \right) = \int_0^\infty  {{e^{ - \frac{{\left( {1 + {\gamma _{\sf th}}} \right){\gamma _{\sf th}}}}{{{\rho _1}{\rho _2}x}}}}{e^{ - x}}{x^{k - N}}}{}_2{F_1}\left( {M + 1,N - m + 1;N - m + 2; - \frac{{{\rho _I}{\gamma _{\sf th}}}}{{{\rho _1}}}\left( {1 + \frac{{{\gamma _{\sf th}} + 1}}{{{\rho _2}x}}} \right)} \right)dx.
\end{align}
\hrule
\end{figure*}
%\begin{multline}
%{{\cal {I}}_1}\left( {{\gamma _{\sf th}}} \right) = \int_0^\infty  {{e^{ - \frac{{\left( {1 + {\gamma _{\sf th}}} \right){\gamma _{\sf th}}}}{{{\rho _1}{\rho _2}x}}}}{e^{ - x}}{x^{k - N}}}\times\\{}_2{F_1}\left( {M + 1,N - m + 1;N - m + 2; - \frac{{{\rho _I}{\gamma _{\sf th}}}}{{{\rho _1}}}\left( {1 + \frac{{{\gamma _{\sf th}} + 1}}{{{\rho _2}x}}} \right)} \right)dx.
%\end{multline}

\proof See Appendix \ref{appendix:theorem:5}. \endproof
\end{theorem}

To the best of the authors' knowledge, the integral ${{\cal {I}}_1}$ does not admit a closed-form expression. Nevertheless, this
single integral expression can be efficiently evaluated numerically. Alternatively, we can use the following closed-form lower bound, which is tight across the entire SNR range, and becomes exact in the high SNR regime.

\begin{corollary}\label{coro:2}
In the presence of interferers at the relay, the outage probability of the dual-hop AF relaying system with the MMSE/MRT scheme is lower bounded by
%\begin{multline}\label{MMSE:3}
%P_{\sf out}^{\sf lMMSE}\left( {{\gamma _{\sf th}}} \right) = 1 - \frac{{\Gamma \left( {N,\frac{{{\gamma _{\sf th}}}}{{{\rho _2}}}} \right)}}{{\Gamma \left( N \right)}}\\
%\times\left[ {\frac{{\Gamma \left( {N,\frac{{{\gamma _{\sf th}}}}{{{\rho _1}}}} \right)}}{{\Gamma \left( N \right)}} - \Gamma \left( {M + 1} \right){e^{ - \frac{{{\gamma _{\sf th}}}}{{{\rho _1}}}}}{{\left( {\frac{{{\gamma _{\sf th}}}}{{{\rho _1}}}} \right)}^N}} \right.\times\\
%\left. {\sum\limits_{m = m_1}^N {{}_2{F_1}\left( {M + 1,N - m + 1,N - m + 2, - \frac{{{\rho _I}}}{{{\rho _1}}}{\gamma _{\sf th}}} \right){\cal B}} } \right],
%\end{multline}
\begin{multline}\label{MMSE:3}
P_{\sf out}^{\sf lMMSE} = 1 - \frac{{\Gamma \left( {N,\frac{{{\gamma _{\sf th}}}}{{{\rho _2}}}} \right)}}{{\Gamma \left( N \right)}}\left[ {\frac{{\Gamma \left( {N,\frac{{{\gamma _{\sf th}}}}{{{\rho _1}}}} \right)}}{{\Gamma \left( N \right)}}} \right. \\
- \Gamma \left( {M + 1} \right){e^{ - \frac{{{\gamma _{\sf th}}}}{{{\rho _1}}}}}{{\left( {\frac{{{\gamma _{\sf th}}}}{{{\rho _1}}}} \right)}^N}\sum\limits_{m = m_1}^N{\rho _I^{N - m + 1}}\times\\
\left. {{\frac{{}_2{F_1}\left( {M + 1,N - m + 1;N - m + 2; - \frac{{{\rho _I}}}{{{\rho _1}}}{\gamma _{\sf th}}} \right)}{{\Gamma \left( m \right)\Gamma \left( {N - m + 2} \right)\Gamma \left( {m - N + M} \right)}}} } \right].
\end{multline}
\end{corollary}
\proof See Appendix \ref{appendix:corollary:2}. \endproof

Having obtained the exact outage probability of the MMSE/MRC scheme, we now look into the high SNR regime, and derive simple analytical approximations for the outage probability of the system.
\begin{theorem}\label{theorem:6}
In the high SNR regime, i.e., ${\rho _2} = \mu {\rho _1}$, ${\rho _1} \to \infty $, the outage probability of the dual-hop AF relaying system with the MMSE/MRT scheme can be approximated as
\begin{multline}\label{MMSE:4}
P_{\sf out}^{\sf MMSE} = \left[ {\Gamma \left( {M + 1} \right)\sum\limits_{m = m_1}^N {\cal A}  + } \right.\\
\left. {\left( {1{\rm{ + }}{{\left( {\frac{1}{\mu }} \right)}^N}} \right)\frac{1}{{\Gamma \left( {N + 1} \right)}}} \right]{\left( {\frac{{{\gamma _{\sf th}}}}{{{\rho _1}}}} \right)^N} + { o}\left({\left( {\frac{{{\gamma _{\sf th}}}}{{{\rho _1}}}} \right)^{N+1}}\right),
\end{multline}
where ${\cal A} = {\frac{{\rho _I^{N - m + 1}}}{{\Gamma \left( m \right)\Gamma \left( {N - m + 2} \right)\Gamma \left( {m - N + M} \right)}}}$.
\end{theorem}
\proof
When ${\rho _2} = \mu {\rho _1}$, ${\rho _1} \to \infty$, we have ${e^{ - \frac{\gamma_{\sf th}}{{{\rho _1}}}}} \to 1$, ${}_2{F_1}\left( {a,b;c; - \frac{{{\rho _I}}}{{{\rho _1}}}{\gamma _{\sf th}}} \right) \to 1$. Together with these observations and with the help of the asymptotic expansion of the incomplete gamma function \cite[Eq. (8.354.2)]{Tables}, we can easily obtain the desired result from {\textit{Corollary}} \ref{coro:2} after some simple algebraic manipulations.
\endproof

Theorem \ref{theorem:6} indicates that the MMSE/MRT scheme achieves a diversity order of $N$, the same as the MRC/MRT scheme. Since the MMSE/MRT scheme in general needs more CSI compared with the MRC/MRT scheme, it is natural to ask whether the MMSE/MRT scheme always achieves a strictly better outage performance. This is indeed the case in the high SNR regime, as shown in the following corollary.
\begin{corollary}\label{coro:3}
In the high SNR regime, the outage probability of dual-hop AF relaying system with the MMSE/MRT scheme is strictly smaller than that of the MRC/MRT scheme.
\end{corollary}
\proof See Appendix \ref{app:coro:3}.
\endproof

%In summary, {\textit{Theorem}} \ref{theorem:4} implies that the ZF/MRT scheme only achieves diversity order of $N-M$, {\textit{Theorem}} \ref{theorem:2} and {\textit{Theorem}} \ref{theorem:6} reveal that both the MRC/MRT and the MMSE/MRT scheme can achieve full diversity gain, i.e., diversity order of $N$. See Table \ref{summary:table} for a summary of the performance comparison among the three proposed schemes. In fact, under the assumption that the relay processing matrix ${\bf{W}}$  has the form ${\bf{W}} = \omega \frac{{{\bf{h}}_2^{\bf{\dag }}}}{{\left| {{{\bf{h}}_2}} \right|}}{{\bf{w}}_1}$, it can be easily proved that the MRC/MRT scheme is a scheme to maximize the molecular of the end-to-end expression in (\ref{SINR:1}) for it match both the first and the second hop channel, and the ZF/MRT scheme is a scheme to minimize the denominator in expression (\ref{SINR:1}) because it force the interference item to zero, while the MMSE/MRT scheme is a scheme to maximize the whole expression (\ref{SINR:1}) since the MMSE combining vector is a kind of optimum combining vector \cite{C.A.Baird}. Therefore, we can predict that the MMSE/MRT scheme will have the best performance among the three proposed schemes while the MRC/MRT scheme and the ZF/MRT scheme will have their own advantage when compared with each other. The subsequent Numerical results in the next section can verify our prediction.

\subsection{Large $N$ Analysis}
In this section, we look into the large $N$ regime with fixed $M$, and examine the asymptotic behavior of the three proposed schemes. We have the following key result.
\begin{theorem}\label{theorem:7}
When $N \to \infty $, the end-to-end SINR of the ZF/MRT and the MMSE/MRT schemes converges to
\begin{align}\label{LN:1}
\gamma^{\infty}=\frac{{{\rho _2}{\left\|{\bf h} _2\right\|_F^2}{\rho _1}{\left\|{\bf h} _1\right\|_F^2}}}{{{\rho _2}{\left\|{\bf h} _2\right\|_F^2} + {\rho _1}{\left\|{\bf h} _1\right\|_F^2} + 1}},
\end{align}
and the corresponding outage probability is given by
\begin{multline}
P _{\sf out}^{\infty} =  1 - \frac{{2{e^{ - \frac{{{\gamma _{\sf th}}}}{{{\rho _1}}} - \frac{{{\gamma _{\sf th}}}}{{{\rho _2}}}}}}}{{\Gamma \left( N \right)}}{\sum\limits_{m = 0}^{N - 1} {\left( {\frac{{{\gamma _{\sf th}}}}{{{\rho _1}}}} \right)} ^m}\frac{1}{{m!}}\sum\limits_{j = 0}^m {m\choose j}\\
 \times{\left( {\frac{1}{{{\rho _2}}}} \right)^{m - j}}\sum\limits_{k = 0}^{N + j - 1} {N+j-1\choose k} {\left( {\frac{{{\gamma _{\sf th}}}}{{{\rho _2}}}} \right)^{N + j - k - 1}}\\
 \times{\left( {\frac{{\left( {1 + {\gamma _{\sf th}}} \right){\gamma _{\sf th}}}}{{{\rho _1}{\rho _2}}}} \right)^{\frac{{k - m + 1}}{2}}}{{K} _{k - m + 1}}\left( {2\sqrt {\frac{{\left( {1 + {\gamma _{\sf th}}} \right){\gamma _{\sf th}}}}{{{\rho _1}{\rho _2}}}} } \right).
\end{multline}
\end{theorem}
\proof See Appendix \ref{appendix:theorem:7}. \endproof

%However, for MRC/MRT scheme, observing the SINR expression in (\ref{PA:1}), we have proven in Appendix \ref{appendix:theorem:1} that the interference term $U_1 = \frac{{\sum\limits_{i = 1}^M {{{\left| {{\bf{h}}_{1}^{\bf{\dag }}{{\bf{h}}_{{Ii}}}} \right|}^2}{\rho_{Ii}}} }}{{{{\left\| {{{\bf{h}}_{1}}} \right\|}_F^2}}}$ is a random variable which have nothing to do with $N$. Hence, no matter how large $N$ is, it dose not affect the distribution of interference term.
{A close observation reveals that the asymptotic SINR $\gamma^{\infty}$ presented in Theorem \ref{theorem:7} is equivalent to the end-to-end SNR of the same dual-hop AF relaying system without CCI at the relay, which suggests that, when $N$ is large, CCI at the relay has no impact on the ZF/MRT and the MMSE/MRT schemes. However, this is not the case for the MRC/MRT scheme. Let us make a careful scrutiny of the interference term $U_1 = \frac{{\sum\limits_{i = 1}^M {{{\left| {{\bf{h}}_{1}^{\bf{\dag }}{{\bf{h}}_{{Ii}}}} \right|}^2}{\rho_{Ii}}} }}{{{{\left\| {{{\bf{h}}_{1}}} \right\|}_F^2}}}$ in (\ref{PA:1}). It can be readily observed that $U_1$ is a hyper-exponential random variable which is independent of $N$. Hence, for the MRC/MRT scheme, the effect of CCI persists even in the large $N$ regime. This implies that in the presence of CCI, if the relay is equipped with a large number of antennas, adopting linear diversity combining schemes with superior interference suppression capability such as ZF/MRT and MMSE/MRT is preferred over the simple MRC/MRT scheme.}

%the outage performance of the ZF/MRT and MMSE/MRT schemes will definitely outperform that of MRC/MRT scheme.

%The underlying reason for such phenomena is that the channel vectors become orthogonal to each other when $N\to \infty$. Hence, as long as the transmission is aligned with the source-relay channel, the effect of CCI disappears.
% However, care should also be taken when interpreting the above result. While it is true that ZF/MRT and MMSE/MRT schemes will achieve the same performance when $N\to \infty$, we demonstrate numerically that the rates at which they approach the same limit differ significantly.

\subsection{Comparison of the Proposed Schemes}
{We now provide a more concrete comparison for the three different schemes studied. In the preceding analysis, the channel state information (CSI) requirement to perform relay precoding was not explicitly revealed. In practice, the acquisition of CSI involves additional feedback overhead, which must be considered in the design of wireless systems. On the other hand, if a large amount of CSI is available at the transmitting node, more sophisticated transmission schemes could be designed to improve the transmission efficiency and to achieve a better performance. Therefore, in order to make a fair comparison among the three different schemes, the CSI requirement of each individual scheme must be characterized. Table I gives a comparison of the MRC/MRT, ZF/MRT and MMSE/MRT schemes.}
\begin{table*}
\centering
\caption{Comparison of the MRC/MRT, ZF/MRT and MMSE/MRT Schemes}
\begin{tabular}{|p{4cm}|c|c|c|}
\hline
{} & MRC/MRT & ZF/MRT & MMSE/MRT \\
\hline
CSI requirement & ${\bf h}_1$ and ${\bf h}_2$ & ${\bf h}_1$, ${\bf h}_2$ and ${\bf H}_I$  & ${\bf h}_1$, ${\bf h}_2$, ${\bf H}_I$, and $N_0$  \\
\hline
Antenna number requirement & None & $N > M$ & None \\
\hline
Diversity order & $N$ & $N - M$ & $N$ \\
\hline
Impact of interference power & reduce the array gain & no impact & reduce the array gain \\
\hline
Decay rate of outage probability vs. $N$ & slow & fast & fast \\
\hline
\end{tabular}
\label{summary:table}
\end{table*}

\section{Numerical Results and Discussion}
In this section, we present numerical results to validate the analytical expressions derived in Section III. Note, the integral expressions presented in Theorem 1 and Theorem 5 are evaluated numerically with the build-in functions in Matlab, i.e., the ``quad'' command, and we choose the default absolute error tolerance value $1.0\times10^{-6}$ to control the accuracy of the numerical integration. In all simulations, we set ${\gamma _{\sf th}} = 0\;{\mbox{dB}}$, ${\rho _{Ii}} = 0\;{\mbox{dB}}, \forall i=1, ... , M$, $\mu = 1$, and all results are obtained with $10^8$ runs.

\begin{figure}[ht]
\centering
\includegraphics[scale=0.5]{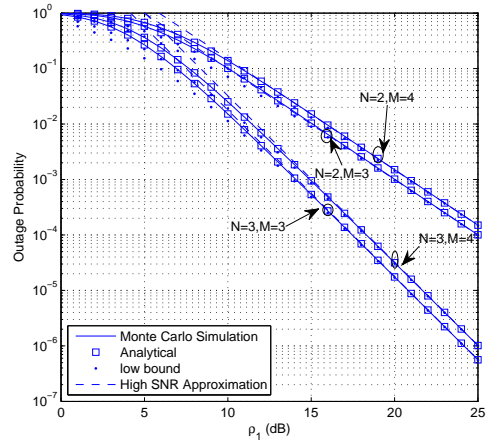}
\caption{Outage probability of the MRC/MRT relaying system with different $M$ and $N$.}\label{fig:fig2}
\end{figure}

Fig. \ref{fig:fig2} shows the outage probability of the dual-hop AF relaying system with the MRC/MRT scheme for different $M$ and $N$. As illustrated, the analytical results are in exact agreement with the Monte Carlo simulations, which demonstrates the correctness of the analytical expression given in (7). Also, the proposed lower bound is sufficiently tight across the entire SNR range of interest, and becomes almost exact in the high SNR regime (i.e., $\rho_1\geq 15\;\mbox{dB}$), while the high SNR approximation works quite well even at moderate SNR values (i.e., ${\rho _1} = 15\;\mbox{dB}$). In addition, we observe that increasing $N$ reduces the outage probability by improving the diversity order of the system, while increasing $M$ degrades the outage performance by reducing the array gain of the system.

\begin{figure}[ht]
\centering
\includegraphics[scale=0.5]{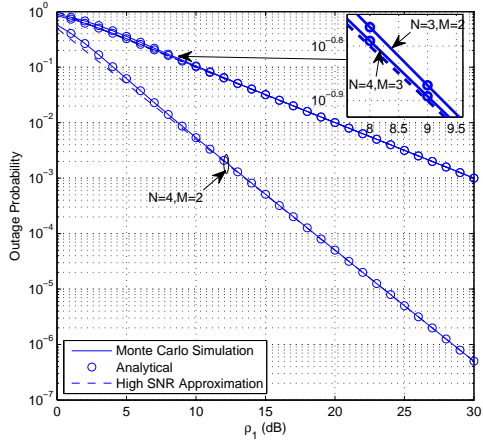}
\caption{Outage probability of the ZF/MRT relaying system with different $M$ and $N$.}\label{fig:fig3}
\end{figure}

Fig. \ref{fig:fig3} illustrates the outage probability of the dual-hop AF relaying system with the ZF/MRT scheme for different $M$ and $N$. It is observed that, for fixed $M$, increasing the antenna number $N$ yields a significant outage improvement, since the achievable diversity order of the system is $N-M$. Moreover, comparing the curves associated with $N=4$, $M=3$ and $N=3$, $M=2$, we observe that, when $N-M$ is fixed, the outage probability difference between different $M$, $N$ pairs is almost negligible, and disappears in the high SNR regime, as shown in Theorem 4.

\begin{figure}[ht]
\centering
\includegraphics[scale=0.5]{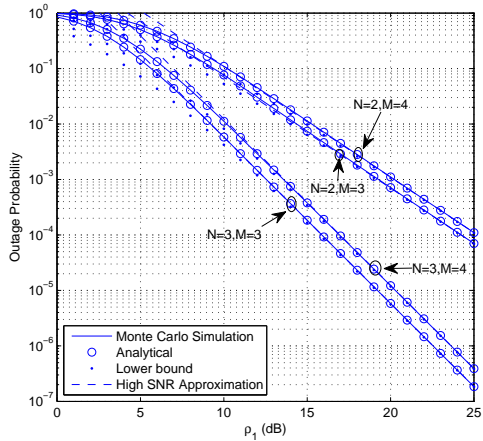}
\caption{Outage probability of the MMSE/MRT relaying system with different $M$ and $N$.}\label{fig:fig4}
\end{figure}

\begin{figure}[ht]
\centering
\includegraphics[scale=0.5]{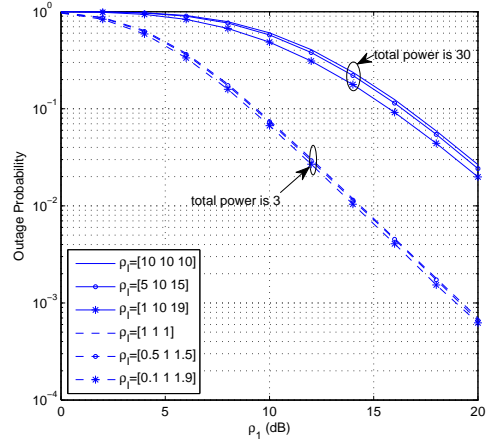}
\caption{Impact of the received interference power distribution on the outage performance for the MMSE/MRT scheme with $N=2$ and $M=3$.}\label{fig:fig5}
\end{figure}

Fig. \ref{fig:fig4} examines the outage probability of the MMSE/MRT scheme for different $M$ and $N$. It can be readily observed that the analytical curves are in perfect agreement with the Monte Carlo simulation results, and the proposed lower bound and the high SNR approximation are sufficiently tight. In addition, similar to the MRC/MRT scheme, we see that the MMSE/MRT scheme achieves a diversity order of $N$.

%As previously mentioned, to make the analysis tractable, we assume the interferers power are equal in the performance analysis of MMSE/MRT scheme.
%Hence, it is necessary to see what happen in the case with unequal-power interferers, and whether the case with equal-power interferers can be a good enough approximation of the case with unequal-power interferers for a given total interference power?

Fig. \ref{fig:fig5} shows the outage probability of the MMSE/MRT scheme when the equal interference power assumption is no longer valid. As we can clearly observe, for a given total interference power, the gap between the analytical results and Monte Carlo simulations is sufficiently small, especially for the low interference power case where the difference is almost negligible. In addition, we see that the curves associated with the equal interference power case have the worst outage performance. Hence, our analytical results could be used to serve as an tight outage upper bound in case of arbitrary interference power profiles.

\begin{figure}[ht]
\centering
\includegraphics[scale=0.5]{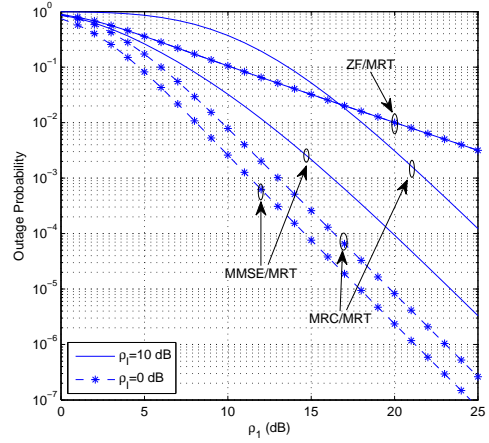}
\caption{Outage probability of MRC/MRT, ZF/MRT and MMSE/MRT schemes with $N = 3$, $M = 2$ and different $\rho_I$.}\label{fig:fig6}
\end{figure}

Fig. \ref{fig:fig6} compares the outage performance of the three relay precoding schemes under different cases of interference power, i.e., weak interference $\rho_I=0 \mbox{ dB}$ and strong interference $\rho_I=30\mbox{ dB}$. We observe the intuitive result that, the outage performance of the MRC/MRT and the MMSE/MRT schemes degrades when the interference power becomes stronger, while the outage performance of the ZF/MRT scheme remains the same regardless of the interference power levels. Comparing different curves, we see that the MMSE/MRT scheme always attains the best outage performance, and the ZF/MRT scheme outperforms the MRC/MRT scheme at the low SNR regime, especially when the interference power is large.

%\begin{figure}[ht]
%\centering
%\includegraphics[scale=0.6]{figure7.eps}
%\caption{Outage probability comparison of MRC/MRT, ZF/MRT and MMSE/MRT schemes with parameter $N = 3$, $M = 2$, $\rho_I=10 \mbox{ dB}$.}\label{fig:fig7}
%\end{figure}

\begin{figure}[ht]
\centering
\includegraphics[scale=0.5]{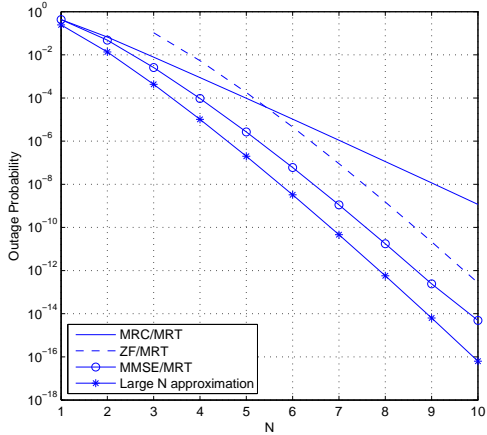}
\caption{Outage comparison of the MRC/MRT, ZF/MRT and MMSE/MRT schemes with $\rho_1=\rho_2=10 \mbox{ dB}$, $\rho_I=0\mbox{ dB}$, $M=2$ and different $N$.}\label{fig:fig7}
\end{figure}

Fig. \ref{fig:fig7} illustrates the impact of relay antenna number on the outage performance of the proposed schemes with fixed source and relay transmit power. We observe that the MMSE/MRT scheme always has the best outage performance, while the MRC/MRT scheme outperforms the ZF/MRT scheme when $N$ is small, and becomes inferior to the ZF/MRT scheme when $N$ is large. Moreover, the outage decay rate of the ZF/MRT and MMSE/MRT schemes is almost the same, which is higher than that of the MRC/MRT scheme. In other words, the minimum required antenna number $N$ to achieve a certain outage probability is smaller for the ZF/MRT and MMSE/MRT schemes compared with the MRC/MRT scheme.

\section{Conclusion}
We investigated the outage performance of a dual-hop AF multiple antenna relaying system employing the MRC/MRT, ZF/MRT and MMSE/MRT schemes in the presence of CCI. Exact analytical expressions for the outage probability of all three schemes were derived, which provide a fast and efficient means of evaluating the performance of the system. Moreover, simple and informative high SNR outage approximations were presented, which enable the characterization of the impact of key parameters, such as relay antenna number $N$, number of interferers $M$ and interference power on the outage performance of the system. Our findings suggest that both the MRC/MRT and MMSE/MRT schemes achieve the full diversity order of $N$, while the ZF/MRT scheme achieves a diversity order of $N-M$. In addition, the MMSE/MRT scheme always attains the best outage performance, and the ZF/MRT scheme outperforms the MRC/MRT scheme in the low SNR regime, while the opposite holds in the high SNR regime. {Finally, we have shown that, in the asymptotically large $N$ regime, perfect interference cancellation can be achieved by using the ZF/MRT and the MMSE/MRT schemes.}

\appendices
\section{Proof for the MRC/MRT Scheme}
\subsection{Proof of Theorem \ref{theorem:1}}\label{appendix:theorem:1}
We start by expressing the end-to-end SINR given in (\ref{PA:1}) as
\begin{align}
\gamma_{\sf MRC}  = \frac{{{y_1}{y_2}{\rho _1}{\rho _2}}}{{\left( {{y_2}{\rho _2} + 1} \right)({U_1} + 1) + {y_1}{\rho _1}}},
\end{align}
where ${y_1} = {{\left\| {{{\bf{h}}_1}} \right\|}_F^2}$, ${y_2} = {{\left\| {{{\bf{h}}_2}} \right\|}_F^2}$, ${U_1} = \sum\limits_{i = 1}^M {{y_{Ii}}{\rho_{Ii}}}$ with ${y_{Ii}} = \frac{{{{\left| {{\bf{h}}_{1}^\dag {{\bf{h}}_{{Ii}}}} \right|}^2}}}{{{{\left\| {{{\bf{h}}_{1}}} \right\|}_F^2}}}$. It is easy to observe that $y_1$ and $y_2$ are i.i.d. random variables with the p.d.f.
\begin{align}
{f_{{y_i}}}(x) = \frac{{{x^{N - 1}}}}{{(N - 1)!}}{e^{ - x}}.
\end{align}
Also, according to \cite{H.Ding}, $y_{Ii}$, $i=1,\dots,M$, are i.i.d. exponential random variables with unit variance. Then, random variable $U_1$ follows the hyper-exponential distribution with p.d.f.
\begin{align}
{f_{{U_1}}}(x) = \sum\limits_{i = 1}^{\rho ({\bf{D}})} {\sum\limits_{j = 1}^{{\tau _i}({\bf{D}})} {{\chi _{i,j}}({\bf{D}})\frac{{\rho _{I\left\langle i \right\rangle }^{ - j}}}{{(j - 1)!}}{x^{j - 1}}{e^{ - \frac{x}{{{\rho _{I\left\langle i \right\rangle }}}}}}} }.
\end{align}
The outage probability of the system can be computed by
\begin{align}
{P_{\sf out}^{\sf MRC}}&= {\mathop{\rm Prob}\nolimits} \left(\frac{{{y_1}{y_2}{\rho _1}{\rho _2}}}{{\left( {{y_2}{\rho _2} + 1} \right)({U_1} + 1) + {y_1}{\rho _1}}} \le {\gamma _{\sf th}}\right) \notag\\
&= {\mathop{\rm Prob}\nolimits} \left( {{y_1}\frac{{{y_2} - \frac{{{\gamma _{\sf th}}}}{{{\rho _2}}}}}{{{y_2} + \frac{1}{{{\rho _2}}}}} \le \frac{{{\gamma _{\sf th}}}}{{{\rho _1}}}\left( {{U_1} + 1} \right)} \right).
\end{align}
To this end, invoking the result presented in \cite[Lemma 3]{C.Zhong1}, we can obtain the following outage probability expression conditioned on $U_1$,
\begin{multline}
P_{\sf out}^{\sf MRC} = 1 - \frac{{2{e^{ - \frac{{{\gamma _{\sf th}}}}{{{\rho _1}}}\left( {{U_1} + 1} \right) - \frac{{{\gamma _{\sf th}}}}{{{\rho _2}}}}}}}{{\Gamma \left( N \right)}}\sum\limits_{m = 0}^{N - 1}{{\left( {\frac{{{\gamma _{\sf th}}}}{{{\rho _1}}}\left( {{U_1} + 1} \right)} \right)}^m}\frac{1}{{m!}}\\
\times\sum\limits_{j = 0}^m {{m\choose j}{{\left( {\frac{1}{{{\rho _2}}}} \right)}^{m - j}}\sum\limits_{k = 0}^{N + j - 1} {N+j-1\choose k} }  \\
\times{\left( {\frac{{{\gamma _{\sf th}}}}{{{\rho _2}}}} \right)^{N + j - 1 - k}}{{\cal B}^{\frac{{k - m + 1}}{2}}}{K_{k - m + 1}}\left( {2\sqrt{\cal B}} \right),
\end{multline}
where ${\cal B} = \frac{{\left( {{\gamma _{\sf th}} + 1} \right){\gamma _{\sf th}}}}{{{\rho _1}{\rho _2}}}\left( {{U_1} + 1} \right)$

The desired result can be obtained by averaging over $U_1$, along with some simple basic algebraic manipulations.

\subsection{Proof of Corollary \ref{coro:1}}\label{appendix:corollary:1}
The end-to-end SINR can be upper bounded by
\begin{align}
&\gamma_{\sf MRC}  = \notag\\
&\frac{{{y_1}{y_2}{\rho _1}{\rho _2}}}{{\left( {{y_2}{\rho _2} + 1} \right)({U_1} + 1) + {y_1}{\rho _1}}} \le \min \left( {\frac{{{y_1}{\rho _1}}}{{{U_1} + 1}},{y_2}{\rho _2}} \right).
\end{align}
Since ${y_1}$, ${y_2}$, ${U_1}$ are independent random variables, the outage probability of the system can be lower bounded by
\begin{align}\label{align:a4}
P_{\sf out}^{\sf lMRC} = 1 - {\mathop{\rm Prob}\nolimits} \left( {\frac{{{y_1}{\rho _1}}}{{{U_1} + 1}} \ge {\gamma _{\sf th}}} \right){\mathop{\rm Prob}\nolimits} \left( {{y_2}{\rho _2} \ge {\gamma _{\sf th}}} \right).
\end{align}
Conditioned on $U_1$, ${\mathop{\rm Prob}\nolimits}\left( {\frac{{{y_1} {\rho _1}}}{{{U_1}  + 1}} \ge {\gamma _{\sf th}}} \right)$ can be computed as
\begin{align}
{\mathop{\rm Prob}\nolimits}\left( {\frac{{{y_1} {\rho _1}}}{{{U_1}  + 1}} \ge {\gamma _{\sf th}}} \right) &= 1 - {\mathop{\rm Prob}\nolimits}\left[ {{y_1}  < \frac{{{\gamma _{\sf th}}}}{{{\rho _1}}}\left( {{U_1}  + 1} \right)} \right]\notag\\
&= \frac{{\Gamma \left( {N,\frac{{{U_1}  + 1}}{{{\rho _1}}}{\gamma _{\sf th}}} \right)}}{{\Gamma \left( N \right)}}.
\end{align}
%We know ${\mathop{\rm Prob}\nolimits}\left( {\frac{{{y_1} {\rho _1}}}{{{U_1}  + 1}} \ge {\gamma _{th}}} \right) = 1 - {\mathop{\rm Prob}\nolimits}\left[ {{y_1}  < \frac{{{\gamma _{th}}}}{{{\rho _1}}}\left( {{U_1}  + 1} \right)} \right]$, hence we obtain this probability conditioned on  ${U_1}$
%\begin{align}\label{align:a9}
%{\mathop{\rm Prob}\nolimits}\left( {\frac{{{y_1} {\rho _1}}}{{{U_1}  + 1}} \ge {\gamma _{th}}} \right) = {e^{ - \frac{{{U_1}  + 1}}{{{\rho _1}}}{\gamma _{th}}}}\sum\limits_{k = 0}^{N - 1} {\frac{{{{\left( {\frac{{{U_1}  + 1}}{{{\rho _1}}}{\gamma _{th}}} \right)}^k}}}{{k!}}}  = \frac{{\Gamma \left( {N,\frac{{{U_1}  + 1}}{{{\rho _1}}}{\gamma _{th}}} \right)}}{{\Gamma \left( N \right)}}.
%\end{align}
The next step is to average over the distribution of ${U_1}$. After some algebraic manipulations, we arrive at
\begin{multline}\label{multline:a5}
{\mathop{\rm Prob}\nolimits}\left( {\frac{{{y_1} {\rho _1}}}{{{U_1}  + 1}} \ge {\gamma _{\sf th}}} \right) = {e^{ - \frac{{{\gamma _{\sf th}}}}{{{\rho _1}}}}}\sum\limits_{k = 0}^{N - 1} {{\left( {\frac{{{\gamma _{\sf th}}}}{{{\rho _1}}}} \right)}^k}\frac{1}{{k!}}\sum\limits_{l = 0}^k {k\choose l}\\
\sum\limits_{i = 1}^{\rho ({\bf {D}} )} {\sum\limits_{j = 1}^{{\tau _i}({\bf {D}} )} {{\chi _{i,j}}({\bf {D}} )\frac{{\Gamma \left( {j + l} \right)}}{{\Gamma \left( j \right)}}} }\rho _{I\left\langle i \right\rangle}^l{{\left( {\frac{{{\rho _1}}}{{{\rho _1} + {\rho _{I\left\langle i \right\rangle}}{\gamma _{\sf th}}}}} \right)}^{j + l}}.
\end{multline}
Now, we look at the second part, ${\mathop{\rm Prob}\nolimits}\left( {{y_2} {\rho _2} \ge {\gamma _{\sf th}}} \right)$, which can be computed as
\begin{align}\label{align:a6}
&{\mathop{\rm Prob}\nolimits}\left( {{y_2} {\rho _2} \ge {\gamma _{\sf th}}} \right) = {\mathop{\rm Prob}\nolimits}\left( {{y_2}  \ge \frac{{{\gamma _{\sf th}}}}{{{\rho _2}}}} \right) \notag\\
&= {e^{ - \frac{{{\gamma _{\sf th}}}}{{{\rho _2}}}}}\sum\limits_{m = 0}^{N - 1} {\frac{{{{\left( {\frac{{{\gamma _{\sf th}}}}{{{\rho _2}}}} \right)}^m}}}{{m!}}}  = \frac{{\Gamma \left( {N,\frac{{{\gamma _{\sf th}}}}{{{\rho _2}}}} \right)}}{{\Gamma \left( N \right)}}.
\end{align}
To this end, substituting (\ref{multline:a5}) and (\ref{align:a6}) into (\ref{align:a4}) yields the desired result.

\subsection{Proof of Theorem \ref{theorem:2}}\label{appendix:theorem:2}
Starting from (\ref{align:a4}), conditioned on ${U_1}$, the outage probability of the MRC/MRT scheme can be lower bounded by
\begin{align}
P_{\sf out}^{\sf lMRC} = 1 - \frac{{\Gamma \left( {N,\frac{{{\gamma _{\sf th}}}}{{{\rho _2}}}} \right)}}{{\Gamma \left( N \right)}}\frac{{\Gamma \left( {N,\frac{{{U_1}  + 1}}{{{\rho _1}}}{\gamma _{\sf th}}} \right)}}{{\Gamma \left( N \right)}}.
\end{align}
Then, invoking the asymptotic expansion of incomplete gamma function \cite[Eq. (8.354.2)]{Tables}, we have
\begin{align}
&\frac{{\Gamma \left( {N,\frac{{{\gamma _{\sf th}}}}{{{\rho _2}}}} \right)}}{{\Gamma \left( N \right)}} = 1 - \frac{{\left( {\frac{{{\gamma _{\sf th}}}}{{{\rho _1}}}} \right)^N}}{{\Gamma \left( {N + 1} \right)}\mu ^N} + { o}\left({\left( {\frac{{{\gamma _{\sf th}}}}{{{\rho _1}}}} \right)^{N+1}}\right).
\end{align}
%\begin{align}
%&\frac{{\Gamma \left( {N,\frac{{{\gamma _{\sf th}}}}{{{\rho _2}}}} \right)}}{{\Gamma \left( N \right)}} =
%1 - \frac{1}{{N\Gamma \left( N \right)}} {{{\left( {\frac{{{\gamma _{\sf th}}}}{{{\rho _2}}}} \right)}^N}} + { o}\left({\left( {\frac{{{\gamma _{\sf th}}}}{{{\rho _2}}}} \right)^{N+1}}\right)\notag\\
% &= 1 - \frac{1}{{\Gamma \left( {N + 1} \right)}}\left( {\frac{1}{{{\mu ^N}}}} \right){\left( {\frac{{{\gamma _{\sf th}}}}{{{\rho _1}}}} \right)^N} + { o}\left({\left( {\frac{{{\gamma _{\sf th}}}}{{{\rho _1}}}} \right)^{N+1}}\right).
%\end{align}
Similarly, we get
\begin{align}\label{align:a10}
&\frac{{\Gamma \left( {N,\frac{{{U_1}  + 1}}{{{\rho _1}}}{\gamma _{\sf th}}} \right)}}{{\Gamma \left( N \right)}} =\notag\\
 &1 - \frac{1}{{\Gamma \left( {N + 1} \right)}}{\left( {{U_1}  + 1} \right)^N}{\left( {\frac{{{\gamma _{\sf th}}}}{{{\rho _1}}}} \right)^N} + { o}\left({\left( {\frac{{{\gamma _{\sf th}}}}{{{\rho _1}}}} \right)^{N+1}}\right).
\end{align}
Hence, the outage lower bound conditioned on ${U_1}$ can be approximated as
%\begin{align}\label{align:a11}
%P_{\sf out}^{\sf lMRC} = \frac{{\left( {\frac{{{\gamma _{\sf th}}}}{{{\rho _1}}}} \right)^N}}{{\Gamma \left( {N + 1} \right)}}\left[ {\frac{1}{{{\mu ^N}}} + {{\left( {{U_1}  + 1} \right)}^N}} \right] + { o}\left({\left( {\frac{{{\gamma _{\sf th}}}}{{{\rho _1}}}} \right)^{N+1}}\right).
%\end{align}
\begin{align}\label{align:a11}
&P_{\sf out}^{\sf lMRC} = \notag\\
&\frac{{\left( {\frac{{{\gamma _{\sf th}}}}{{{\rho _1}}}} \right)^N}}{{\Gamma \left( {N + 1} \right)}}\left[ {\frac{1}{{{\mu ^N}}} + {{\left( {{U_1}  + 1} \right)}^N}} \right] + { o}\left({\left( {\frac{{{\gamma _{\sf th}}}}{{{\rho _1}}}} \right)^{N+1}}\right).
\end{align}
To this end, the remaining task is to compute the expectation of $\left( {1 + {U_1} } \right)^N$. Applying the binomial expansion,
\begin{align}
{\left( {1 + {U_1} } \right)^N} = \sum\limits_{k = 0}^N {N\choose k} {{U_1} ^k},
\end{align}
and averaging over $U_1$, we get
\begin{align}
&{E_{{U_1} }}\left\{\left( {1 + {U_1} } \right)^N\right\} = \sum\limits_{k = 0}^N {N\choose k} \sum\limits_{i = 1}^{\rho ({\bf{D}})} \sum\limits_{j = 1}^{{\tau _i}({\bf{D}})} {{\chi _{i,j}}({\bf{D}})\frac{{{{{\rho _{I\left\langle i \right\rangle }} }^{ - j}}}}{{(j - 1)!}}}\notag\\
&\times{{\int_0^\infty  {{{{x} }^{k + j - 1}}{e^{ - \frac{{{x} }}{{{\rho _{I\left\langle i \right\rangle }} }}}}} d{x} } }\notag\\
  &= \sum\limits_{k = 0}^N {N\choose k} \sum\limits_{i = 1}^{\rho ({\bf{D}})} {\sum\limits_{j = 1}^{{\tau _i}({\bf{D}})} {{\chi _{i,j}}({\bf{D}})\frac{{\Gamma \left( {k + j} \right)}}{{\Gamma \left( j \right)}}\rho _{I\left\langle i \right\rangle}^k} }.\label{align:a12}
\end{align}
%\begin{align}
%{f_{{U_1}}}(x) = \sum\limits_{i = 1}^{\rho ({\bf{D}})} {\sum\limits_{j = 1}^{{\tau _i}({\bf{D}})} {{\chi _{i,j}}({\bf{D}})\frac{{\rho _{I\left\langle i \right\rangle }^{ - j}}}{{(j - 1)!}}{x^{j - 1}}{e^{ - \frac{x}{{{\rho _{I\left\langle i \right\rangle }}}}}}} }
%\end{align}
Substituting (\ref{align:a12}) into (\ref{align:a11}) yields the desired result.

\section{Proof for the ZF/MRT Scheme}
\subsection{Proof of Proposition \ref{prop:11}}\label{app:prop:11}
Substituting ${{\bf{w}}_1}{{\bf{H}}_I} = {\bf{0}}$ and $\left| {{\bf{w}}_1} \right| = 1$ into (\ref{SINR:2}), we have
\begin{align}
\gamma_{\sf ZF}=\frac{{{\omega ^2}{{\left\| {{{\bf{h}}_2}} \right\|}_F^2}{{\left| {\;{{{\bf{w}}_1}}{{\bf{h}}_1}} \right|}^2}{\rho_1}}}{{{\omega ^2}{{\left\| {{{\bf{h}}_2}} \right\|}_F^2} + 1}}.
\end{align}
It should be noted that the power constraint constant $\omega$ is also dependent on the combining vector $\mathbf{w}_1$ via the following relationship
\begin{align}
{\omega ^2} = \frac{{{\rho_2}}}{{\left| {{\bf{w}}_1{{\bf{h}}_1}} \right|^2{\rho_1} + 1}}.
\end{align}
Hence, we have
\begin{align}
\gamma_{\sf ZF}=\frac{{{{\left\| {{{\bf{h}}_2}} \right\|}_F^2}{\rho_2}{{\left| {\;{{{\bf{w}}_1}}{{\bf{h}}_1}} \right|}^2}{\rho_1}}}{{{{\left\| {{{\bf{h}}_2}} \right\|}_F^2}{\rho_2} + {{(\left| {{\bf{w}}_1{{\bf{h}}_1}} \right|^2{\rho_1} + 1)}}}}.
\end{align}
Now, it is easy to show that $\gamma_{\sf ZF}$ is an increasing function with respect to $\left| {{\bf{w}}_1{{\bf{h}}_1}} \right|^2$. Therefore, the original optimization can be alternatively formulated as
\begin{align}\label{ZF:3}
\begin{array}{l}
{{\bf{w}}_1} = \arg \;\mathop {\max }\limits_{{\bf{w}}_1} \;\;{\left| {\;{{\bf{w}}_1}{{\bf{h}}_1}} \right|^2}\\
{\text{s.t.}}\;\;{{\bf{w}}_1}{{\bf{H}}_I} = {\bf{0}}\;\;\& \;\;\left\| {{\bf{w}}_1} \right\|_F = 1.
\end{array}
\end{align}
To this end, the desired result can be obtained by following the similar lines as in the proof of \cite[Proposition 1]{Z.Ding}.

\subsection{Proof of Theorem \ref{theorem:4}}\label{appendix:theorem:4}
In the high SNR regime, the end-to-end SINR $\gamma_{\sf ZF}$ can be tightly bounded by
\begin{align}
\gamma_{\sf ZF} = \frac{{{y_2}{y_3}{\rho _1}{\rho _2}}}{{{y_2}{\rho _2} + {y_3}{\rho _1} + 1}} \le \min \left( {{y_3}{\rho _1},{y_2}{\rho _2}} \right).
\end{align}
Hence, the outage probability of the system can be tightly lower bounded by
\begin{align}
{P_{\sf out}^{\sf ZF}} \ge {\mathop{\rm Prob}\nolimits}\left( {\min \left( {{y_3}{\rho _1},{y_2}{\rho _2}} \right) \le {\gamma _{\sf th}}} \right).
\end{align}
Due to the independence of random variables $y_2$ and $y_3$, the outage lower bound can be written as
\begin{align}
{P_{\sf out}^{\sf ZF}} & = 1 - {\mathop{\rm Prob}\nolimits}\left( {{y_3} \ge \frac{{{\gamma _{\sf th}}}}{{{\rho _1}}}} \right){\mathop{\rm Prob}\nolimits}\left( {{y_2} \ge \frac{{{\gamma _{\sf th}}}}{{{\rho _2}}}} \right)\notag\\
& = 1 - \frac{{\Gamma \left( {N - M,\frac{{{\gamma _{\sf th}}}}{{{\rho _1}}}} \right)}}{{\Gamma \left( {N - M} \right)}}\frac{{\Gamma \left( {N,\frac{{{\gamma _{\sf th}}}}{{{\rho _2}}}} \right)}}{{\Gamma \left( N \right)}}.
\end{align}
Invoking the asymptotic expansion of incomplete gamma function \cite[Eq. (8.354.2)]{Tables}, we have
\begin{align}\label{appendix:ZF1}
&{P _{\sf out}^{\sf ZF}} = \frac{{\left( {\frac{{{\gamma _{\sf th}}}}{{{\rho _1}}}} \right)^N}}{{\Gamma \left( {N + 1} \right)}}{\left( {\frac{1}{\mu }} \right)^N} + \frac{{\left( {\frac{{{\gamma _{\sf th}}}}{{{\rho _1}}}} \right)^{N - M}}}{{\Gamma \left( {N - M + 1} \right)}}\notag\\
&+ { o}\left({\left( {\frac{{{\gamma _{\sf th}}}}{{{\rho _1}}}} \right)^{N-M+1}}\right).
\end{align}
It is obvious that the first item in (\ref{appendix:ZF1}) is negligible when compared with the second item in (\ref{appendix:ZF1}). Therefore, the desired result can be obtained after some simple algebraic manipulations.

\section{Proof for the MMSE/MRT Scheme}
\subsection{Proof of Proposition \ref{proposition:2}}\label{appendix:proposition:2}
The random variable $Z$ can be alternatively expressed as
\begin{align}\label{MMSE:a1}
Z = \frac{{{\rho_1}}}{{{\rho_I}}}\frac{{{{\bf{w}}_1}{{\bf{h}}_1}{\bf{h}}_1^\dag {\bf{w}}_1^\dag }}{{{{\bf{w}}_1}\left( {\sum\limits_{i = 1}^M {{{\bf{h}}_{Ii}}{\bf{h}}_{Ii}^\dag } } \right){\bf{w}}_1^\dag  + {{\bf{w}}_1}\left( {\frac{{{1}}}{{{\rho_I}}}{\bf{I}}} \right){\bf{w}}_1^\dag }},
\end{align}
where ${\bf{I}}$ denotes the unit matrix.

Now define ${\bf{R}} = {{\bf{H}}_I}{\bf{H}}_I^\dag  + \frac{{{1}}}{{{\rho_I}}}{\bf{I}}$, then the MMSE combining vector ${\bf w}_1$ can be written by
\begin{align}
{{\bf{w}}_1} = {\bf{h}}_1^\dag {\left( {{{\bf{h}}_1}{\bf{h}}_1^\dag  + {\bf{R}}} \right)^{ - 1}}.
\end{align}
Hence, we have
\begin{align}\label{MMSE:a2}
Z = \frac{{{\rho_1}}}{{{\rho_I}}}\frac{{{\bf{h}}_1^\dag {{\left( {{{\bf{h}}_1}{\bf{h}}_1^\dag  + {\bf{R}}} \right)}^{ - 1}}{{\bf{h}}_1}{\bf{h}}_1^\dag {{\left( {{{\bf{h}}_1}{\bf{h}}_1^\dag  + {\bf{R}}} \right)}^{ - 1}}{{\bf{h}}_1}}}{{{\bf{h}}_1^\dag {{\left( {{{\bf{h}}_1}{\bf{h}}_1^\dag  + {\bf{R}}} \right)}^{ - 1}}{\bf{R}}{{\left( {{{\bf{h}}_1}{\bf{h}}_1^\dag  + {\bf{R}}} \right)}^{ - 1}}{{\bf{h}}_1}}}.
\end{align}
Applying the well-known matrix inversion lemma, (\ref{MMSE:a2}) can be simplified as
\begin{align}\label{eqn:equz}
Z = \frac{{{\rho_1}}}{{{\rho_I}}}{\bf{h}}_1^\dag {{\bf{R}}^{ - 1}}{{\bf{h}}_1}.
\end{align}
To this end, invoking the results presented in \cite[Eq.(11)]{H.Gao}, the c.d.f. of $Z$ can be expressed as
%\begin{align}
%{{\mathop{\rm F}\nolimits} _{{Z_1}}}\left( z \right) = 1 - {e^{ - \frac{z}{{{\rho_I }}}}}\sum\limits_{m = 1}^N {\frac{{{A_m}\left( z \right)}}{{\left( {m - 1} \right)!}}{{\left( {\frac{z}{{{\rho_I}}}} \right)}^{m - 1}}},
%\end{align}
%where ${A_m}\left( z \right) = \left\{ {\begin{array}{*{20}{c}}
%{1\;\;\;\;\;\;\;\;\;\;\;\;\;\;\;\;\;\;\;\;\;\;\;N \ge M + m}\\
%{\frac{{1 + \sum\limits_{i = 1}^{N - m} {{M\choose i}{z^i}} }}{{{{\left( {1 + z} \right)}^M}}}\;\;\;\;\;\;N < M + m}
%\end{array}} \right.$.
%
%Hence, the c.d.f. of $Z$ can be computed as ${{\mathop{\rm F}\nolimits} _Z}\left( z \right) = {\mathop{\rm Prob}\nolimits}\left( {\frac{{{P}}}{{{P_I}}}{Z_1} < z} \right) = {{\mathop{\rm F}\nolimits} _{{Z_1}}}\left( {\frac{{{P_I}}}{{{P}}}z} \right)$, and the c.d.f. of $Z$ can be shown as
\begin{align}\label{MMSE:a3}
{{F} _Z}\left( z \right) = 1 - {e^{ - \frac{z}{{{\rho _1}}}}}\sum\limits_{m = 1}^N {\frac{{{A_m}\left( z \right)}}{{\left( {m - 1} \right)!}}{{\left( {\frac{z}{{{\rho _1}}}} \right)}^{m - 1}}},
\end{align}
where
\begin{equation}
{A_m}\left( z \right) = \left\{ \begin{array}{cc}
1 & N \ge M + m,\\
\frac{{1 + \sum\limits_{i = 1}^{N - m} {{M\choose i}{{\left( {\frac{{{\rho _I}}}{{{\rho _1}}}z} \right)}^i}} }}{{{{\left( {1 + \frac{{{\rho _I}}}{{{\rho _1}}}z} \right)}^M}}} & N < M + m.\\
\end{array} \right.
\end{equation}
To obtain a unified expression for the c.d.f. of $Z$, we find it convenient to give a separate treatment for the following two cases: 1) $N\ge M$, and 2) $N\leq M$.
\subsubsection{$N\ge M$}
Noticing that
\begin{align}
\frac{{1 + \sum\limits_{i = 1}^{N - m} {{M\choose i}{{\left( {\frac{{{\rho _I}}}{{{\rho _1}}}z} \right)}^i}} }}{{{{\left( {1 + \frac{{{\rho _I}}}{{{\rho _1}}}z} \right)}^M}}} = 1 - \frac{{\sum\limits_{i = N - m + 1}^M {{M\choose i}{{\left( {\frac{{{\rho _I}}}{{{\rho _1}}}{\rm{z}}} \right)}^i}} }}{{{{\left( {1 + \frac{{{\rho _I}}}{{{\rho _1}}}z} \right)}^M}}},
\end{align}
the c.d.f. of $Z$ can be written as (\ref{MMSE:a11}) shown on the top of the next page,
\begin{figure*}
\begin{align}\label{MMSE:a11}
{{\mathop{\rm F}\nolimits} _Z}\left( z \right) = 1 - {e^{ - \frac{z}{{{\rho _1}}}}} {\sum\limits_{m = 1}^{N - M} \frac{1}{{\left( {m - 1} \right)!}}{\left( {\frac{z}{{{\rho _1}}}} \right)}^{m - 1}}\sum\limits_{m = N - M + 1}^N {\frac{{e^{ - \frac{z}{{{\rho _1}}}}}}{{\left( {m - 1} \right)!}}}{{\left( {\frac{z}{{{\rho _1}}}} \right)}^{m - 1}}\left( {1 - \frac{{\sum\limits_{i = N - m + 1}^M {{M\choose i}{{\left( {\frac{{{\rho _I}}}{{{\rho _1}}}{\rm{z}}} \right)}^i}} }}{{{{\left( {1 + \frac{{{\rho _I}}}{{{\rho _1}}}z} \right)}^M}}}} \right),
\end{align}
\hrule
\end{figure*}
%\begin{align}\label{MMSE:a11}
%&{{\mathop{\rm F}\nolimits} _Z}\left( z \right) = \notag\\
%&1 - {e^{ - \frac{z}{{{\rho _1}}}}} {\sum\limits_{m = 1}^{N - M} \frac{1}{{\left( {m - 1} \right)!}}{\left( {\frac{z}{{{\rho _1}}}} \right)}^{m - 1}}\sum\limits_{m = N - M + 1}^N {\frac{{e^{ - \frac{z}{{{\rho _1}}}}}}{{\left( {m - 1} \right)!}}}\notag\\
%&\times{{\left( {\frac{z}{{{\rho _1}}}} \right)}^{m - 1}}\left( {1 - \frac{{\sum\limits_{i = N - m + 1}^M {{M\choose i}{{\left( {\frac{{{\rho _I}}}{{{\rho _1}}}{\rm{z}}} \right)}^i}} }}{{{{\left( {1 + \frac{{{\rho _I}}}{{{\rho _1}}}z} \right)}^M}}}} \right),
%\end{align}
%\begin{align}
%&{{\mathop{\rm F}\nolimits} _Z}\left( z \right) = 1 - {e^{ - \frac{z}{{{\rho _1}}}}} {\sum\limits_{m = 1}^{N - M} \frac{1}{{\left( {m - 1} \right)!}}{\left( {\frac{z}{{{\rho _1}}}} \right)}^{m - 1}}\times \notag\\
%& \sum\limits_{m = N - M + 1}^N {\frac{{e^{ - \frac{z}{{{\rho _1}}}}}}{{\left( {m - 1} \right)!}}{{\left( {\frac{z}{{{\rho _1}}}} \right)}^{m - 1}}\left( {1 - \frac{{\sum\limits_{i = N - m + 1}^M {{M\choose i}{{\left( {\frac{{{\rho _I}}}{{{\rho _1}}}{\rm{z}}} \right)}^i}} }}{{{{\left( {1 + \frac{{{\rho _I}}}{{{\rho _1}}}z} \right)}^M}}}} \right)},
%\end{align}
which can be further simplified as
\begin{align}
&{{\mathop{\rm F}\nolimits} _Z}\left( z \right) = 1 - {e^{ - \frac{z}{{{\rho _1}}}}}\sum\limits_{m = 1}^N {\frac{1}{{\left( {m - 1} \right)!}}}{{\left( {\frac{z}{{{\rho _1}}}} \right)}^{m - 1}}-\notag\\
&\sum\limits_{m = N - M + 1}^N {\frac{{e^{ - \frac{z}{{{\rho _1}}}}}}{{\left( {m - 1} \right)!}}{{\left( {\frac{z}{{{\rho _1}}}} \right)}^{m - 1}}\underbrace {\frac{{\sum\limits_{i = N - m + 1}^M {{M\choose i}{{\left( {\frac{{{\rho _I}}}{{{\rho _1}}}{\rm{z}}} \right)}^i}} }}{{{{\left( {1 + \frac{{{\rho _I}}}{{{\rho _1}}}z} \right)}^M}}}}_{{S_1}}}.
\end{align}
%Make a comparison between \cite[Eq.(18)]{H.Gao} and \cite[Eq.(16)]{A.Shah}, we can observe an interesting result as follows:
Now, utilizing the following key observation,
\begin{align}\label{MMSE:a4}
&\frac{{\sum\limits_{i = N}^M {{M\choose i}{{\left( {\frac{{{\rho _I}}}{{{\rho _1}}}z} \right)}^i}} }}{{{{\left( {1 + \frac{{{\rho _I}}}{{{\rho _1}}}z} \right)}^M}}} = \frac{{\Gamma \left( {M + 1} \right)}}{{\Gamma \left( {N + 1} \right)\Gamma \left( {M + 1 - N} \right)}}{\left( {\frac{{{\rho _I}}}{{{\rho _1}}}z} \right)^N}\notag\\
&\times{}_2{F_1}\left( {M + 1,N;N + 1; - \frac{{{\rho _I}}}{{{\rho _1}}}z} \right).
\end{align}
$S_1$ can be alternatively expressed as
\begin{align}
&{S_1} = \notag\\
&\frac{{\Gamma \left( {M + 1} \right){}_2{F_1}\left( {M + 1,N - m + 1;N - m + 2; - \frac{{{\rho _I}}}{{{\rho _1}}}z} \right)}}{{\Gamma \left( {N - m + 2} \right)\Gamma \left( {m - N + M} \right){\left( {\frac{{{\rho _I}}}{{{\rho _1}}}{z}} \right)^{-N + m - 1}}}}.
\end{align}
Finally, noticing that
\begin{align}
{e^{ - \frac{z}{{{\rho _1}}}}}\sum\limits_{m = 1}^N {\frac{1}{{\left( {m - 1} \right)!}}{{\left( {\frac{z}{{{\rho _1}}}} \right)}^{m - 1}} = \frac{{\Gamma \left( {N,\frac{z}{{{\rho _1}}}} \right)}}{{\Gamma \left( N \right)}}},
\end{align}
and after some algebraic manipulations, we obtain (\ref{MMSE:a6}) shown on the top of the next page.
\begin{figure*}
\begin{align}\label{MMSE:a6}
{{F} _Z}\left( z \right) = 1 - \frac{{\Gamma \left( {N,\frac{z}{{{\rho _1}}}} \right)}}{{\Gamma \left( N \right)}} + \Gamma \left( {M + 1} \right){e^{ - \frac{z}{{{\rho _1}}}}}{\left( {\frac{z}{{{\rho _1}}}} \right)^N}\sum\limits_{m = N - M + 1}^N {{\rho _I^{N - m + 1}}}\frac{{}_2{F_1}\left( {M + 1,N - m + 1;N - m + 2; - \frac{{{\rho _I}}}{{{\rho _1}}}z} \right)}{{\Gamma \left( m \right)\Gamma \left( {N - m + 2} \right)\Gamma \left( {m - N + M} \right)}}.
\end{align}
\hrule
\end{figure*}
%\begin{multline}\label{MMSE:a6}
%{{F} _Z}\left( z \right) = \\
%1 - \frac{{\Gamma \left( {N,\frac{z}{{{\rho _1}}}} \right)}}{{\Gamma \left( N \right)}} + \Gamma \left( {M + 1} \right){e^{ - \frac{z}{{{\rho _1}}}}}{\left( {\frac{z}{{{\rho _1}}}} \right)^N}\sum\limits_{m = N - M + 1}^N {{\rho _I^{N - m + 1}}}\\
%\times\frac{{}_2{F_1}\left( {M + 1,N - m + 1;N - m + 2; - \frac{{{\rho _I}}}{{{\rho _1}}}z} \right)}{{\Gamma \left( m \right)\Gamma \left( {N - m + 2} \right)\Gamma \left( {m - N + M} \right)}}.
%\end{multline}
\subsubsection{$N \leq M$}
Similarly, the c.d.f. of $Z$ can be alternatively expressed as
\begin{align}\label{MMSE:a7}
&{{\mathop{\rm F}\nolimits} _Z}\left( z \right)
 = 1 - {e^{ - \frac{z}{{{\rho _1}}}}}\sum\limits_{m = 1}^N \frac{1}{{\left( {m - 1} \right)!}}{{\left( {\frac{z}{{{\rho _1}}}} \right)}^{m - 1}}-\notag\\
 & \sum\limits_{m = 1}^N {\frac{{e^{ - \frac{z}{{{\rho _1}}}}}}{{\left( {m - 1} \right)!}}{{\left( {\frac{z}{{{\rho _1}}}} \right)}^{m - 1}}\frac{{\sum\limits_{i = N - m + 1}^M {{M\choose i}{{\left( {\frac{{{\rho _I}}}{{{\rho _1}}}{z}} \right)}^i}} }}{{{{\left( {1 + \frac{{{\rho _I}}}{{{\rho _1}}}z} \right)}^M}}}}.
\end{align}
Then, following the same lines as in the derivation of the $N\ge M$ case, we get (\ref{MMSE:a8}) shown on the top of the next page.
\begin{figure*}
\begin{align}\label{MMSE:a8}
{{F} _Z}\left( z \right) = 1 - \frac{{\Gamma \left( {N,\frac{z}{{{\rho _1}}}} \right)}}{{\Gamma \left( N \right)}} + \Gamma \left( {M + 1} \right){e^{ - \frac{z}{{{\rho _1}}}}}{\left( {\frac{z}{{{\rho _1}}}} \right)^N}\sum\limits_{m = 1}^N {{\rho _I^{N - m + 1}}}\frac{{}_2{F_1}\left( {M + 1,N - m + 1;N - m + 2; - \frac{{{\rho _I}}}{{{\rho _1}}}z} \right)}{{\Gamma \left( m \right)\Gamma \left( {N - m + 2} \right)\Gamma \left( {m - N + M} \right)}}.
\end{align}
\hrule
\end{figure*}
%\begin{multline}\label{MMSE:a8}
%{{F} _Z}\left( z \right) = \\
%1 - \frac{{\Gamma \left( {N,\frac{z}{{{\rho _1}}}} \right)}}{{\Gamma \left( N \right)}} + \Gamma \left( {M + 1} \right){e^{ - \frac{z}{{{\rho _1}}}}}{\left( {\frac{z}{{{\rho _1}}}} \right)^N}\sum\limits_{m = 1}^N {{\rho _I^{N - m + 1}}}\\
%\times\frac{{}_2{F_1}\left( {M + 1,N - m + 1;N - m + 2; - \frac{{{\rho _I}}}{{{\rho _1}}}z} \right)}{{\Gamma \left( m \right)\Gamma \left( {N - m + 2} \right)\Gamma \left( {m - N + M} \right)}}.
%\end{multline}
To this end, the desired result can be obtained by appropriately choosing certain parameters.

\subsection{Proof of Theorem \ref{theorem:5}}\label{appendix:theorem:5}
Starting from the end-to-end SINR presented in (\ref{MMSE:21}), the outage probability of the system can be expressed as
\begin{align}\label{MMSE:a9}
{P_{\sf out}^{\sf MMSE}}& = {\mathop{\rm Prob}\nolimits}\left( {\frac{{{\rho _2}{y_2}Z}}{{\left( {{\rho _2}{y_2} + 1} \right) + Z}} \le {\gamma _{\sf th}}} \right) \notag\\
&= {\mathop{\rm Prob}\nolimits}\left( {Z\frac{{{y_2} - \frac{{{\gamma _{\sf th}}}}{{{\rho _2}}}}}{{{y_2} + \frac{1}{{{\rho _2}}}}} \le {\gamma _{\sf th}}} \right).
\end{align}
Then, applying the method used in \cite[Lemma 3]{C.Zhong1}, and
%\begin{align}
%{P_{\sf out}^{\sf MMSE}} = \int_{\frac{{{\gamma _{\sf th}}}}{{{\rho _2}}}}^\infty  {{{\mathop{\rm F}\nolimits} _z}\left[ {\frac{{\left( {t + \frac{1}{{{\rho _2}}}} \right){\gamma _{\sf th}}}}{{\left( {t - \frac{{{\gamma _{\sf th}}}}{{{\rho _2}}}} \right)}}} \right]{{\mathop{\rm f}\nolimits} _{{y_2}}}\left( t \right)dt}  + \int_0^{\frac{{{\gamma _{\sf th}}}}{{{\rho _2}}}} {\;{{\mathop{\rm f}\nolimits} _{{y_2}}}\left( t \right)dt}.
%\end{align}
utilizing the c.d.f. expression of $Z$ given in Proposition \ref{proposition:2} and the p.d.f of $y_2$, the outage probability of the system is given by
\begin{align}\label{MMSE:a10}
{P_{\sf out}^{\sf MMSE}} &= 1 -  {\cal I}_2+{\left( {\frac{{{\gamma _{\sf th}}}}{{{\rho _1}}}} \right)^N}\frac{{\Gamma \left( {M + 1} \right)}}{{\Gamma \left( N \right)}}\notag\\
&\sum\limits_{m = m_1}^N {\frac{{\rho _I^{N - m + 1}}{\cal I}_3}{{\Gamma \left( m \right)\Gamma \left( {N - m + 2} \right)\Gamma \left( {m - N + M} \right)}}},
\end{align}
where
\begin{align}
{\cal I}_2=\int_{\frac{{{\gamma _{\sf th}}}}{{{\rho _2}}}}^\infty  {\frac{{\Gamma \left( {N,{\cal C}\frac{{{\gamma _{\sf th}}}}{{{\rho _2}}}} \right)}}{{\Gamma \left( N \right)}}\frac{{{t^{N - 1}}}}{{\Gamma \left( N \right)}}{e^{ - t}}dt},
\end{align}
and
\begin{multline}
{\cal I}_3=\int_{\frac{{{\gamma _{\sf th}}}}{{{\rho _2}}}}^\infty  {{e^{ - {\cal C}\frac{{{\gamma _{\sf th}}}}{{{\rho _2}}}}}{e^{ - t}}{t^{N - 1}}{{\cal C}}^N}\\ {}_2{F_1}\left( {M + 1,N - m + 1;N - m + 2; - \frac{{{\rho _I}{\gamma _{\sf th}}}}{{{\rho _1}}}{\cal C}} \right)dt,
\end{multline}
with ${\cal C} = \frac{{t + /\rho _2}}{{t - \gamma _{\sf th}/\rho _2}}$.

To this end, after some tedious algebraic manipulations, solving the integrals ${\cal I}_2$ and ${\cal I}_3$ yields the the desired result.

%we obtainwith the help of \cite[Lemma 3]{C.Zhong1}, the integral ${\cal I}_2$ can be solved as
%\begin{multline}
%{\cal I}_2 = 1 - \frac{{2{e^{ - \frac{{{\gamma _{\sf th}}}}{{{\rho _1}}} - \frac{{{\gamma _{\sf th}}}}{{{\rho _2}}}}}}}{{\Gamma \left( N \right)}}{\sum\limits_{m = 0}^{N - 1} {\left( {\frac{{{\gamma _{\sf th}}}}{{{\rho _1}}}} \right)} ^m}\frac{1}{{m!}}\sum\limits_{j = 0}^m {m\choose j} {\left( {\frac{1}{{{\rho _2}}}} \right)^{m - j}}\sum\limits_{k = 0}^{N + j - 1} {N+j-1\choose k} \\
%{\left( {\frac{{{\gamma _{\sf th}}}}{{{\rho _2}}}} \right)^{N + j - k - 1}}{\left( {\frac{{\left( {1 + {\gamma _{\sf th}}} \right){\gamma _{\sf th}}}}{{{\rho _1}{\rho _2}}}} \right)^{\frac{{k - m + 1}}{2}}}{{K} _{k - m + 1}}\left( {2\sqrt {\frac{{\left( {1 + {\gamma _{\sf th}}} \right){\gamma _{\sf th}}}}{{{\rho _1}{\rho _2}}}} } \right).
%\end{multline}
%
%To compute ${\cal I}_3$, we first make a change of variable $x = t - \frac{{{\gamma _{\sf th}}}}{{{\rho _1}}}$. Then, applying the binomial expansion, and after some tedious algebraic manipulations, we obtain
%\begin{align}
%{{{\cal {I}}}_3} = {e^{ - \frac{{{\gamma _{\sf th}}}}{{{\rho _1}}} - \frac{{{\gamma _{\sf th}}}}{{{\rho _2}}}}}{\sum\limits_{j = 0}^N {{N\choose j}\left( {\frac{1}{{{\rho _2}}}} \right)} ^{N - j}}{\sum\limits_{k = 0}^{N + j - 1} {{N+j-1\choose k}\left( {\frac{{{\gamma _{\sf th}}}}{{{\rho _2}}}} \right)} ^{N + j - 1 - k}}{{{\cal {I}}}_1}\left( {{\gamma _{\sf th}}} \right).
%\end{align}
%To this end, pulling everything together yields the desired result.

\subsection{Proof of Corollary \ref{coro:2}}\label{appendix:corollary:2}
The end-to-end SINR can be upper bounded by
\begin{align}
\gamma_{\sf MMSE}  = \frac{{{\rho _2}{y_2}Z}}{{\left( {{\rho _2}{y_2} + 1} \right) + Z}} \le \min \left( {Z,{\rho _2}{y_2}} \right).
\end{align}
Hence, the outage probability can be lower bounded by
\begin{align}
{P_{\sf out}^{\sf MMSE}} \ge {\mathop{\rm Prob}\nolimits}\left( {\min \left( {Z,{y_2}{\rho _2}} \right) \le {\gamma _{\sf th}}} \right).
\end{align}
Due to the independence of $y_2$ and $Z$, the outage probability lower bound can be computed as
\begin{align}
P_{\sf out}^{\sf lMMSE}\left( {{\gamma _{\sf th}}} \right) = 1 - {\mathop{\rm Prob}\nolimits}\left( {Z \ge {\gamma _{\sf th}}} \right){\mathop{\rm Prob}\nolimits}\left( {{y_2} \ge \frac{{{\gamma _{\sf th}}}}{{{\rho _2}}}} \right).
\end{align}
To this end, invoking the c.d.f. of $Z$ given in Proposition \ref{proposition:2} and the p.d.f of $y_2$, the desired result can be obtained after some simple algebraic manipulations.

\subsection{Proof of Corollary \ref{coro:3}}\label{app:coro:3}
To prove the statement, we only need to show that $A^{\sf MRC}> A^{\sf MMSE}$, where
\begin{align}
A^{\sf MRC}=\frac{1}{{\Gamma \left( {N + 1} \right)}}\left[ {\frac{1}{{{\mu ^N}}} + \sum\limits_{k = 0}^N {N\choose k} \frac{{\Gamma \left( {k + M} \right)}}{{\Gamma \left( M \right)}}\rho _I^k} \right],
\end{align}
and
\begin{multline}
A^{\sf MMSE}= \sum\limits_{m = m_1}^N {\frac{\Gamma \left( {M + 1} \right){\rho _I^{N - m + 1}}}{{\Gamma \left( m \right)\Gamma \left( {N - m + 2} \right)\Gamma \left( {m - N + M} \right)}}}  + \\
\left( {1{\rm{ + }}{{\left( {\frac{1}{\mu }} \right)}^N}} \right)\frac{1}{{\Gamma \left( {N + 1} \right)}}
\end{multline}
A close observation shows that both $A^{\sf MRC}$ and $A^{\sf MMSE}$ have the a common item $\left( {1{\rm{ + }}{{\left( {\frac{1}{\mu }} \right)}^N}} \right)\frac{1}{{\Gamma \left( {N + 1} \right)}}$. Hence, to proof $A^{\sf MRC}> A^{\sf MMSE}$, we only need to show $A_1> A_2$, where
\begin{align}\label{eqn:a1}
A_1= \sum\limits_{k = 1}^N \frac{\Gamma(M+k)\rho _I^k}{\Gamma(N-k+1)\Gamma(k+1)\Gamma(M)},
\end{align}
and
\begin{align}
A_2= \sum\limits_{m = m_1}^N {\frac{\Gamma \left( {M + 1} \right){\rho _I^{N - m + 1}}}{{\Gamma \left( m \right)\Gamma \left( {N - m + 2} \right)\Gamma \left( {m - N + M} \right)}}}.
\end{align}
Define $t=\min(N,M)$, after some tedious algebraic manipulations, $A_2$ can be alternatively expressed as
\begin{align}\label{eqn:a2}
A_2= \sum\limits_{k = 1}^t {\frac{\Gamma \left( {M + 1} \right){\rho _I^{k}}}{{\Gamma(N-k+1)\Gamma \left( k+1 \right)\Gamma \left( {M - k + 1} \right)}}}.
\end{align}
Comparing (\ref{eqn:a2}) and (\ref{eqn:a1}), it is easy to show that $A_1>A_2$, which completes the proof.

\section{Proof of Theorem \ref{theorem:7}}\label{appendix:theorem:7}
In the asymptotic large $N$ regime, the law of large number holds, and we have
\begin{align}\label{align:a13}
\frac{1}{N}{\bf{h}}_1^\dag {{\bf{h}}_{Ii}} = 0,\mbox{ and }\frac{1}{N}{\bf{H}}_I^\dag {{\bf{H}}_I} = {{\bf{I}}_M}.
\end{align}

%For MRC/MRT scheme, substituting (\ref{align:a13}) into (\ref{PA:1}) and simplifying, we have
%\begin{align}
%\gamma_{\sf MRC}  = \frac{{{{\left\| {{{\bf{h}}_2}} \right\|}_F^2}{{\left\| {{{\bf{h}}_1}} \right\|}_F^2}P_s}}{{{{\left\| {{{\bf{h}}_2}} \right\|}_F^2}{N_0} + \frac{{{N_0}}}{{{P_r}}}\left( {{{\left\| {{{\bf{h}}_1}} \right\|}_F^2}P_s + {N_0}} \right)}}=\gamma^{\infty}.
%\end{align}

For the ZF/MRT scheme, starting from (\ref{ZF:5}), and with the help of (\ref{align:a13}), we have
\begin{align}
\left| {{\bf{h}}_1^\dag {\bf{P}}{{\bf{h}}_1}} \right| &= \left| {{\bf{h}}_1^\dag {{\bf{h}}_1} - \frac{1}{N}{\bf{h}}_1^\dag {{\bf{H}}_I}{{\left(\frac{1}{N} {{\bf{H}}_I^\dag {{\bf{H}}_I}} \right)}^{ - 1}}{\bf{H}}_I^\dag {{\bf{h}}_1}} \right|\\
 & = \left| {{\bf{h}}_1^\dag {{\bf{h}}_1} - \frac{1}{N}{\bf{h}}_1^\dag {{\bf{H}}_I}{\bf{H}}_I^\dag {{\bf{h}}_1}} \right| \\
 &= \left| {{\bf{h}}_1^\dag {{\bf{h}}_1} - \frac{1}{N}\sum\limits_{i = 1}^M {{\bf{h}}_1^\dag {{\bf{h}}_{Ii}}{\bf{h}}_{Ii}^\dag {{\bf{h}}_1}} } \right|= \left| {{\bf{h}}_1^\dag {{\bf{h}}_1}} \right|\\
 & = {\left\| {{{\bf{h}}_1}} \right\|_F^2}.\label{align:a15}
\end{align}
Substituting (\ref{align:a15}) into (\ref{ZF:5}), we have
\begin{align}
\gamma_{\sf ZF}  = \frac{{{{\left\| {{{\bf{h}}_2}} \right\|}_F^2}{{\left\| {{{\bf{h}}_1}} \right\|}_F^2}P_s{P_r}}}{{{P_r}{{\left\| {{{\bf{h}}_2}} \right\|}_F^2}{N_0} + {N_0}\left( {{{\left\| {{{\bf{h}}_1}} \right\|}_F^2}P_s + {N_0}} \right)}}=\gamma^{\infty}.
\end{align}

For the MMSE/MRT scheme, with the help of (\ref{align:a13}), and applying the well-known Woodbury matrix identity, we have ${{\bf{R}}^{ - 1}} = \rho_I{\bf{I}} - \frac{\rho_I}{ N} {{\bf{H}}_I}{\bf{H}}_I^\dag.
$
Hence, $Z$ can be expressed as
\begin{align}\label{align:a16}
Z = \frac{\rho_1}{\rho_I}\left(\rho_I{\bf{h}}_1^\dag {{\bf{h}}_1} - \frac{\rho_I}{N}{\bf{h}}_1^\dag {{\bf{H}}_I}{\bf{H}}_I^\dag {{\bf{h}}_1}\right) = \rho_1\left\|\bf{h}_1\right\|_F^2.
\end{align}
Substituting (\ref{align:a16}) into (\ref{MMSE:21}), we have
\begin{align}
\gamma_{\sf MMSE}  = \frac{{\rho_2}{{\left\| {{{\bf{h}}_2}} \right\|}_F^2}\rho_1\left\|\bf{h}_1\right\|_F^2}{{{1+\rho_2}{{\left\| {{{\bf{h}}_2}} \right\|}_F^2}+\rho_1\left\|\bf{h}_1\right\|_F^2}}=\gamma^{\infty}.
\end{align}

Given the asymptotic SINR, the desired outage probability can be computed by following the similar lines as in the proof of Theorem \ref{theorem:3}.

\nocite{*}
\bibliographystyle{IEEE}
\begin{footnotesize}

\end{footnotesize}

\end{document}